\documentclass[nohyperref]{article}

\usepackage[accepted]{icml2022}
\usepackage{tablefootnote}
\usepackage{threeparttable}
\usepackage[utf8]{inputenc} % allow utf-8 input
\usepackage[T1]{fontenc}    % use 8-bit T1 fonts
\usepackage{hyperref}       % hyperlinks
\usepackage{url}            % simple URL typesetting
\usepackage{booktabs}       % professional-quality tables
\usepackage{amsfonts}       % blackboard math symbols
\usepackage{nicefrac}       % compact symbols for 1/2, etc.
\usepackage{microtype}      % microtypography
\usepackage{xcolor}         % colors
\usepackage{standalone}
\usepackage{latexsym}
\usepackage{amsmath}
\usepackage{amssymb}
\usepackage{amsthm}
\usepackage{graphicx}
\usepackage{subcaption}
\usepackage{array}
\usepackage{tabu}
\usepackage{makecell}
\usepackage{subcaption}
\usepackage{paralist}
\usepackage{cases}
\usepackage{diagbox}
\usepackage{enumitem}
\usepackage{soul}
\usepackage{multirow}
\usepackage{verbatim}
\usepackage{tabulary}
\usepackage{booktabs}
\usepackage[mathscr]{euscript}
\usepackage{mathtools}
\usepackage{algorithm}

\usepackage{algpseudocode}
\usepackage{stmaryrd}
\usepackage{tikz-dependency}
\usepackage{subcaption}
\usetikzlibrary{automata,decorations.markings,arrows,positioning,matrix,calc,patterns,angles,quotes,calc}
\usepackage{adjustbox}
\usepackage{tabularx}
\usepackage{xspace}
\usepackage{tabulary}
\usepackage{afterpage}
\usepackage{hyperref}
\usepackage{url}
\usepackage{bm}
\usepackage{color}
\usepackage{graphicx}
\usepackage{slashbox}
\usepackage[toc,page]{appendix}
\usepackage{makecell}
\usepackage{boldline}
\usepackage{wrapfig}
\usepackage[capitalize,noabbrev]{cleveref}
\usepackage[usestackEOL]{stackengine}
\usepackage{tikz}

\usepackage{inconsolata}

\definecolor{orange}{rgb}{1,0.5,0}
\definecolor{mdgreen}{rgb}{0.05,0.6,0.05}
\definecolor{mdblue}{rgb}{0,0,0.7}
\definecolor{dkblue}{rgb}{0,0,0.5}
\definecolor{dkgray}{rgb}{0.3,0.3,0.3}
\definecolor{slate}{rgb}{0.25,0.25,0.4}
\definecolor{gray}{rgb}{0.5,0.5,0.5}
\definecolor{ltgray}{rgb}{0.7,0.7,0.7}
\definecolor{purple}{rgb}{0.7,0,1.0}
\definecolor{lavender}{rgb}{0.65,0.55,1.0}

\definecolor{mypurple}{RGB}{175,48,235}
\definecolor{myblue}{RGB}{46,88,180}
\definecolor{myred}{RGB}{181,68,106}
\definecolor{myyellow}{RGB}{204,143,55}
\theoremstyle{plain}

\theoremstyle{definition}

\theoremstyle{remark}

\usepackage[textsize=tiny]{todonotes}

\renewcommand{\paragraph}[1]{
     \textbf{#1} 
 }

\begin{document}

\twocolumn[
\icmltitle{DS-1000: A Natural and Reliable Benchmark for Data Science Code Generation}



\icmlsetsymbol{equal}{*}
% \icmlsetsymbol{internship}{*}

\begin{icmlauthorlist}
\icmlauthor{Yuhang Lai${}^{*}$}{hku}
\icmlauthor{\enspace Chengxi Li${}^{*}$}{hku}
\icmlauthor{\enspace Yiming Wang${}^{*}$}{hku,pku}
\icmlauthor{\enspace Tianyi Zhang${}^{*}$}{stanford}
\icmlauthor{\enspace Ruiqi Zhong${}^{*}$}{ucb} \\
\icmlauthor{\enspace Luke Zettlemoyer}{uw,fair}
\icmlauthor{\enspace Scott Wen-tau Yih}{fair}
\icmlauthor{\enspace Daniel Fried}{cmu}
\icmlauthor{\enspace Sida Wang}{fair}
\icmlauthor{\enspace Tao Yu}{hku,uw}
\end{icmlauthorlist}

\icmlaffiliation{hku}{The University of Hong Kong}
\icmlaffiliation{pku}{Peking University}
\icmlaffiliation{stanford}{Stanford University}
\icmlaffiliation{ucb}{UC Berkeley}
\icmlaffiliation{uw}{University of Washington}
% \icmlaffiliation{meta}{Meta AI}
\icmlaffiliation{fair}{Meta AI}
\icmlaffiliation{cmu}{Carnegie Mellon University}

\icmlcorrespondingauthor{Tao Yu}{tyu@cs.hku.hk}

\icmlkeywords{Machine Learning, ICML}

\vskip 0.3in
]



 \printAffiliationsAndNotice{\icmlEqualContribution} % otherwise use the standard text.

\newcommand{\ensuretext}[1]{#1}
\newcommand{\marker}[2]{\ensuremath{^{\textsc{#1}}_{\textsc{#2}}}}
\newcommand{\arkcomment}[3]{\ensuretext{\textcolor{#3}{[#1 #2]}}}

\newcommand{\nascomment}[1]{\arkcomment{\marker{NA}{S}}{#1}{blue}}
\newcommand{\tao}[1]{\arkcomment{\marker{T}{Y}}{#1}{orange}}
\newcommand{\rahul}[1]{\arkcomment{\marker{R}{N}}{#1}{magenta}}
\newcommand{\tianyi}[1]{\arkcomment{\marker{T}{Z}}{#1}{green}}
\newcommand{\dfried}[1]{\arkcomment{\marker{D}{F}}{#1}{cyan}}
\newcommand{\chengxi}[1]{\arkcomment{\marker{C}{L}}{#1}{red}}
\newcommand{\wym}[1]{\arkcomment{\marker{Y}{M}}{#1}{myblue}}
\newcommand{\lyh}[1]{\arkcomment{\marker{Y}{L}}{#1}{mypurple}}
\newcommand{\zrq}[1]{\arkcomment{\marker{R}{Z}}{#1}{myyellow}}

\newcommand{\eg}{e.g.,\xspace}
\newcommand{\ie}{i.e.,\xspace}
\newcommand{\dscg}{DS-1000\xspace}
\newcommand{\np}{NumPy\xspace}
\newcommand{\pd}{Pandas\xspace}
\newcommand{\py}{PyTorch\xspace}
\newcommand{\tf}{TensorFlow\xspace}
\newcommand{\scp}{SciPy\xspace}
\newcommand{\sk}{Scikit-learn\xspace}
\newcommand{\plt}{Plotting\xspace}
\newcommand{\mpl}{Matplotlib\xspace}
\newcommand{\SO}{StackOverflow\xspace}
\newcommand{\github}{GitHub\xspace}
\newcommand{\npHundred}{\texttt{numpy-100}\xspace}
\newcommand{\codexTwo}{Codex-002\xspace}
\newcommand{\codexOne}{Codex-001\xspace}
\newcommand{\codexCush}{Codex-Cushman\xspace}
\newcommand{\codegen}{CodeGen\xspace}
\newcommand{\incoder}{InCoder\xspace}

\def\code#1{\texttt{#1}}
\newcommand{\mbr}[1]{MBR-\textsc{Exec}}

\begin{abstract}
We introduce \dscg, a code generation benchmark with a thousand data science problems spanning seven Python libraries, such as \code{\np} and  \code{\pd}.  
Compared to prior works, \dscg incorporates three core features. 
First, our problems reflect diverse, realistic, and practical use cases since we collected them from \SO.
Second, our automatic evaluation is highly specific (reliable) -- across all \codexTwo-predicted solutions that our evaluation accept, only 1.8\% of them are incorrect;
we achieve this with multi-criteria metrics, checking both functional correctness by running test cases and surface-form constraints by restricting API usages or keywords.
Finally, we proactively defend against memorization by slightly modifying our problems to be different from the original \SO source; consequently, models cannot answer them correctly by memorizing the solutions from pre-training.
The current best public system (\codexTwo) achieves 43.3\% accuracy, leaving ample room for improvement.
We release our benchmark at \url{https://ds1000-code-gen.github.io}.

\end{abstract}

\section{Introduction}\label{intro}
Data science is important in many areas \citep{Education-Data-Science, Medicine-Data-Science, Climate-Data-Science}, but requires programming proficiency in specialized libraries, thus posing substantial barriers to lay users.
Fortunately, these barriers could potentially be reduced by pre-trained code generation models:
for example, Codex~\citep{Codex-HumanEval} can complete small Python snippets with non-trivial accuracy and AlphaCode~\citep{alphacode} can tackle difficult competitive programming problems.
We anticipate that these barriers will diminish if the community can make solid progress in applying these models to data science problems. 

However, we currently lack a benchmark that 1) focuses on everyday data science applications, 2) includes naturalistic intents and contexts, and 3) has a reliable execution-based evaluation metric.
Most of the existing datasets with reliable test cases \citep{hendrycksapps2021, Codex-HumanEval} focus on competition or interview-style programming problems;
they measure algorithmic understanding but do not target real-world usage.
Also, as represented by \eg user problems on \SO, users' data science coding problems usually have diverse contexts including their incorrect code, error messages, and input-output examples, which cannot be found in most prior data science relevant code generation benchmarks~\citep{yin2018mining,hendrycksapps2021,chandel2022training,Codex-HumanEval}.
Moreover, most of these benchmarks solely rely on surface-form metrics such as BLEU or CodeBLEU~\citep{yin2018mining, agashe2019juice, chen2021plotcoder}.
These metrics diverge from the programmer's intent, increasingly so as model capability improves~\citep{zhong-etal-2020-semantic}.
To our knowledge, no existing benchmarks contain both naturally occurring problems with diverse contexts and reliable evaluation metrics. 
\begin{figure*}[t!]
    \centering
    \includegraphics[width=0.8\textwidth]{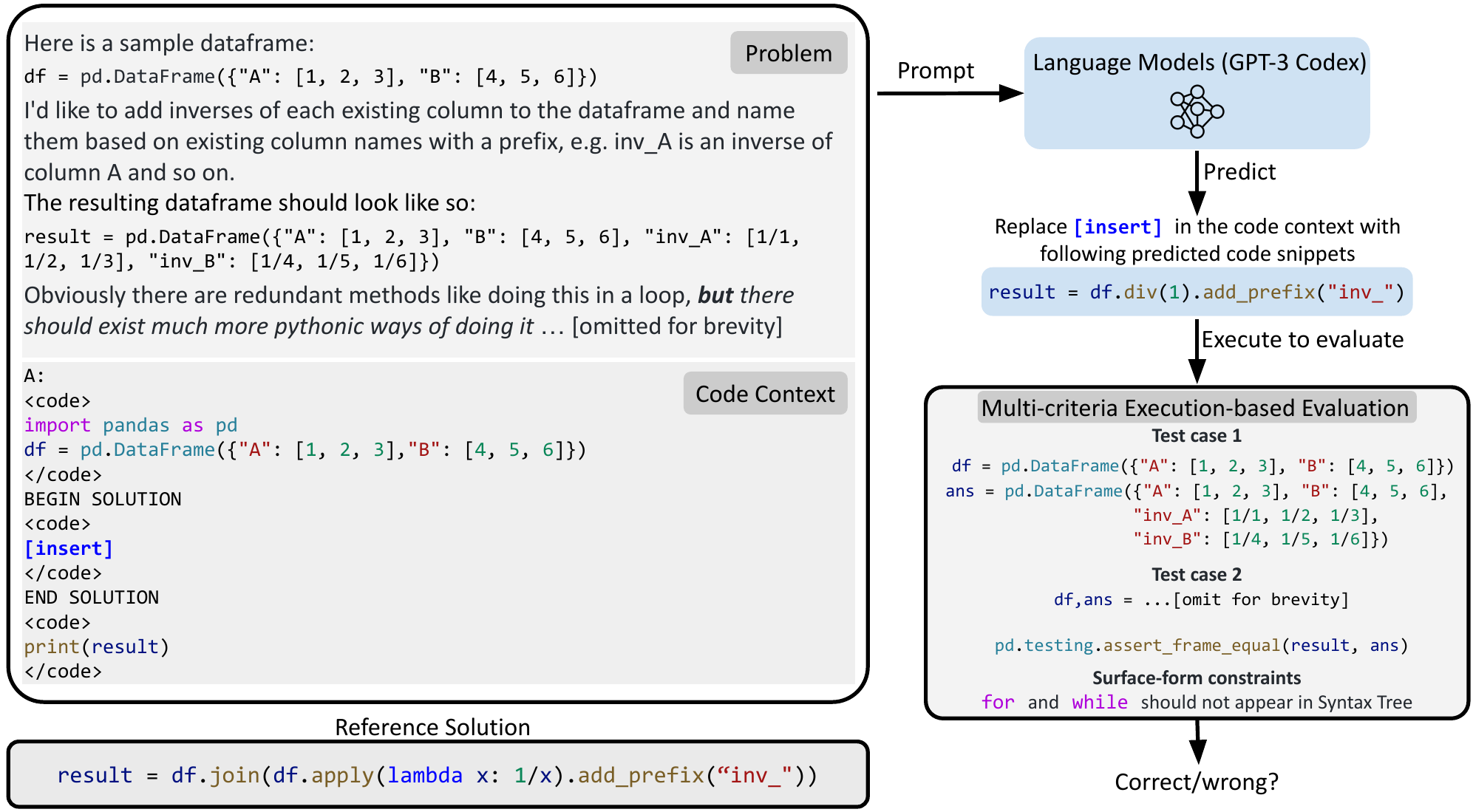}
    \caption{
    An example problem in \dscg. 
    The model needs to fill in the code into ``[insert]'' in the prompt on the left;
    the code will then be executed to pass the multi-criteria automatic evaluation, which includes the test cases and the surface-form constraints;
    a reference solution is provided at the bottom left.
    }
    \label{fig:example-datapoint}
    \vspace{-0.3cm}
\end{figure*}

To fill this gap, we introduce \dscg, a benchmark with a thousand problems covering seven widely-used Python data science libraries: \code{\np}, \code{\pd}, \code{\tf}, \code{\py}, \code{\scp}, \code{\sk}, and \code{\mpl}.
We highlight three core features of \dscg: 1) it contains realistic problems with diverse contexts, 2) it implements reliable multi-criteria execution-based evaluation metrics, and 3) it proactively defends against memorization.
We outline how we achieved each of them below.

First, we collected naturally occurring problems from \SO, manually scored their representativeness and usefulness, and curated a subset of them to create our benchmark.
While inputs in existing code generation datasets are either highly structured (problems or code context) or restricted in scope, our natural problems are diverse in content and format.
For example, users might search for more efficient code implementations (\Cref{fig:example-datapoint}), provide incorrect code with an error message and ask for bug fixes (\Cref{fig:example-pytorch-41}), inquire about specific API usage (\Cref{fig:example-sklearn-60}), or ask for code that implements functionality they specify with input-output examples (\Cref{fig:example-datapoint}).
These problems better reflect real-world applications and open up new modeling challenges, which have been understudied in existing code generation benchmarks.

Second, it is challenging to evaluate program solutions to natural and diverse problems reliably.
Unlike competition-style problems, natural problems might lack executable contexts and test cases, allow multiple solutions, depend on external libraries, etc.
To address these challenges, 
five of the authors of this paper, all proficient in data science and experts in Python, hand-adapted the original problems by writing executable code contexts, rewriting problems to be specific enough to be testable, and implementing automatic multi-criteria execution-based evaluation using carefully written and reviewed test cases and constraints that check functional correctness and surface-form constraints.
On program solutions predicted by \codexTwo, we find that only 1.8\% of the predicted programs passing our evaluation are incorrect (false discovery rate), indicating that our evaluation is reliable.

Third, one potential concern for adapting public problems is that the models might simply memorize the corresponding solution during pre-training time~\citep{274574}.
We show in \Cref{sec:perturb} that this can indeed happen:
while Codex achieves 72.5\% accuracy on the popular \npHundred dataset, the accuracy drastically drops to 23.6\% after perturbing them without increasing their difficulty.
Therefore, while building \dscg, we proactively took measures against memorization by perturbing each problem.

\Cref{fig:example-datapoint} shows an example \dscg problem, its reference solution, and an expert-written automatic multi-criteria evaluation.
To answer the problem, the model needs to fill in the solution; to pass our automatic evaluation, it needs to 1) return the correct output and 2) avoid inefficient implementations that use for-loops.

We use \dscg to evaluate several popular code generation models, including Codex \citep{Codex-HumanEval}, \codegen \citep{codegen}, and \incoder \citep{Incoder}. 
We found model performance ranges from 7.4\% to 43.3\%,
with \codexTwo model being the best.
This implies that these models have the potential to reduce the barrier for data analysis, yet still have large room for improvement.
\section{Benchmark Construction} \label{sec:pipeline}
\begin{figure*}[t]
    \centering
    \includegraphics[width=0.8\textwidth]{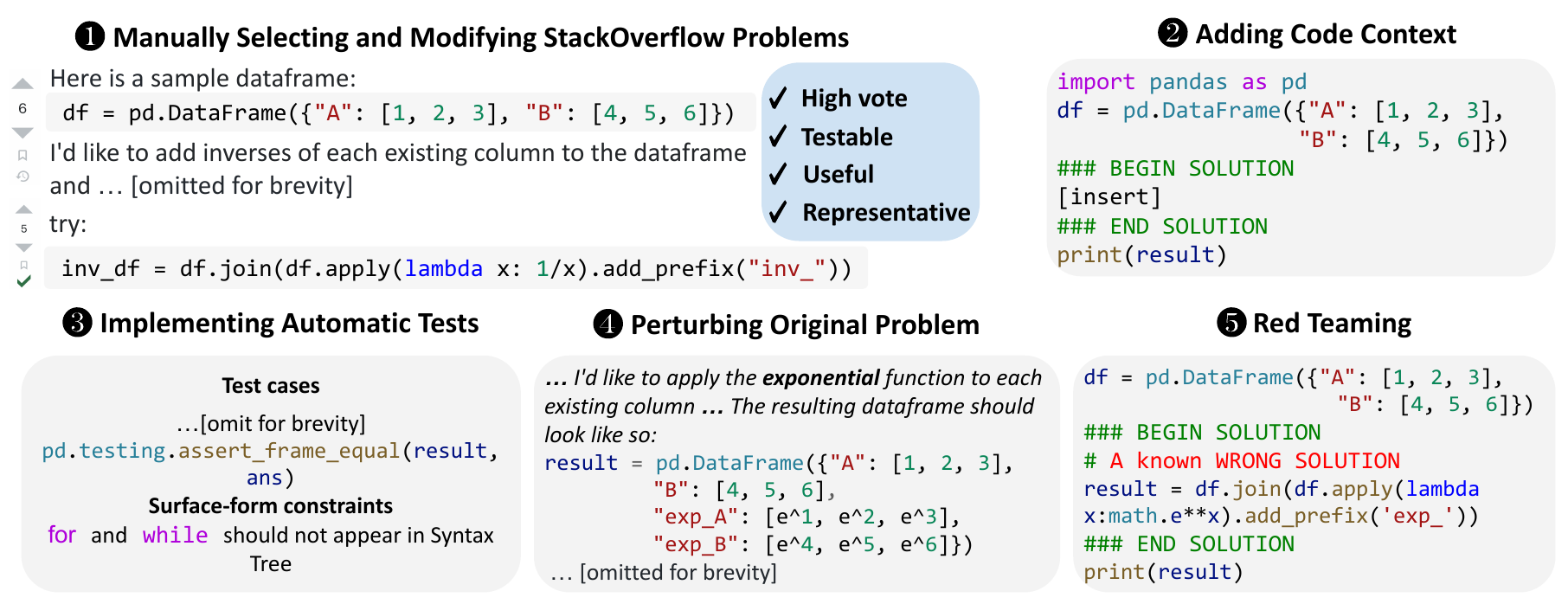}
    \caption{The pipeline for building \dscg. See the start of \Cref{sec:pipeline} for a detailed description.}
    \label{fig:pipeline}
\end{figure*}

Our pipeline for building \dscg contains five stages, illustrated in \Cref{fig:pipeline} and described below.
1) We scraped and selected high-quality problems from \SO (\Cref{sec:selection}).
2) We rewrote the problem and the reference solution so that the problem is unambiguous and the reference solution is executable.(\Cref{sec:rewrite})
3) We implemented a multi-criteria automatic evaluation for each problem, which includes test cases and surface-form constraints (\Cref{sec:mce}). 
4) We performed a pilot study which shows that Codex can answer problems by memorizing the pre-training corpus, and proactively took measures to prevent this by perturbing the problems and their reference solutions in \dscg (\Cref{sec:perturb}).
5) We improved our multi-criteria evaluation by requiring it to reject a small set of sample predictions that we considered incorrect via manual review (\Cref{sec:qa}), and then calculated the false discovery rate of our metric on a larger set of sample predictions.
To reliably carry out this data collection procedure, five authors who are computer science students and familiar with data science spent a total of about 1200 hours constructing \dscg (including steps from problem selection to quality review).

\subsection{Problem Selection} \label{sec:selection}
\paragraph{Sourcing Popular \SO Problems.} To obtain natural and high-quality problems, we scraped data from \SO under each library tag (e.g., ``\code{\np}''). 
To select popular problems, we first removed duplicates and selected problems with at least 1 vote, 1000 views, that had an accepted answer. 
Next, we ranked problems based on votes and views and calibrated these statistics based on the time a problem was created since older problems naturally have more views and votes.
We refer readers to Appendix~\ref{sec:problem-selection-appendix} for more details.
Among the filtered problems, we randomly sampled an initial pool containing 4500 problems (1000 for \code{\np}, \code{\pd}, and \code{\mpl}, 500 for \code{\sk} and \code{\scp}, 250 for \code{\tf}, and 250 for \code{\py}).

\paragraph{Filtering Suitable Problems.} 
To select problems from the above pool for our benchmark, our annotators scored each problem according to the following rubric: whether a problem
a) contains input-output examples in the problem, 
b) is difficult to predict the solution for models according to the annotators' judgment, 
c) is practically useful, 
d) has a clear description, 
and e) is feasible to evaluate the solution.
We aggregated these scores, reranked the candidate problems,  and incorporated the top-ranked ones to create \dscg. 
We ended up with 451 unique \SO problems.
More than half of the original \SO problems were filtered out because they ask for an explanation for an algorithm or general content (see \Cref{sec:problem-selection-appendix}).

\paragraph{Controlling Library Version.} 
Data science libraries are continuously evolving.
As a result, the semantics of the problem is determined not only by the language description but also by the software environment (e.g., library version).
For example, the same code snippet, \code{tf.math.reciprocal(A)}, is only valid in the newer version of \code{\tf}.
We fixed the evaluation environment to include the latest versions of libraries that can be installed with Python 3.7.10 and present the detailed documentation in \Cref{sec:problem-selection-appendix}.

\subsection{Rewriting Problems and Reference Solutions} \label{sec:rewrite}

\paragraph{Creating Executable Context.} 
To implement an execution-based evaluation for each natural language problem, we needed to write an executable context.
We first added package imports and defined the variables described in the problem.
For example, in \Cref{fig:pipeline}, we imported the \code{\pd} package and created the dataframe described in the problem as part of the context.
Second, we needed to specify the desired behavior of the target program to be predicted.
For example, in \Cref{fig:pipeline}, a code generation model can infer from the context that the resulting dataframe should be named as \code{result}, rather than \code{output}.

\begin{table*}[t]
\centering
\renewcommand\arraystretch{1.2}
\resizebox{\linewidth}{!}{
\begin{tabular}{c|l|l}
\hline
Perturbation     & Categories                                               & Example    \\ \hline
{\multirow{3}{*}{Surface}} & Convert to completing function     & \Cref{fig:example-Function}, change format of code context\\
 & Paraphrase the description of the problem           & \Cref{fig:example-A1}, express the same problem in different words    \\
 & Change the example input and output                  & \Cref{fig:example-A2}, replace this example with a longer one    \\ \hline
{\multirow{4}{*}{Semantic}} 
 & Replace keywords with analogy words                  & \Cref{fig:example-A3}, replace ``inv'' with ``exp''    \\
 & Change the required index                            & \Cref{fig:example-A4}, need the specified rows and columns    \\
 & Reverse the order of the list, string or dataframe   & \Cref{fig:example-A5}, reverse the needed string    \\
 & Change the type of the required result               & \Cref{fig:example-A6}, change the \code{DataFrame} to a \code{Series}   \\ \hline
 {\multirow{2}{*}{Difficult Rewrite}} 
 & Combining several surface and semantic perturbations & \Cref{fig:example-P1}, change examples and replace ``highest'' with ``lowest''     \\ 
 & Digging more perturbations that increase the difficulty   & \Cref{fig:example-P2}, hypothesis testing     \\ \hline
\end{tabular}
}
\caption{
The perturbation categories along with examples. ``Surface'' perturbations do not change
the reference solution, while ``Semantic'' perturbations do.
}
\label{tab:example of perturbation}
\vspace{-10pt}
\end{table*}

\paragraph{Rewriting \mpl Problems.}
Many \code{\mpl} problems on \SO clarify their problems with example figures, which, however, cannot be encoded by current pre-trained code models.
Therefore, we rewrote the \SO problems in symbols (i.e., code and text) and adopted a different format from other libraries (see \Cref{fig:example-plotting}). 

\paragraph{Collecting Reference Solutions.}
Finally, we obtained the reference solution for each problem from multiple high-vote replies, edited all reference solutions to be executable given the context we provided, and fixed errors whenever we noticed them (e.g., \Cref{fig:example-pandas-115}).
Even though we did not use the reference solution in \dscg for evaluation, we provided them in \dscg to facilitate future research.

\subsection{Implementing Multi-Criteria Evaluations} \label{sec:mce}

Our automatic evaluation is multi-criteria, checking both functional correctness and surface-form constraints. 

\paragraph{Functional Correctness.}
To evaluate functional correctness, we constructed test cases by converting the input-output examples provided in the \SO problem;
then the expert annotators manually wrote additional test cases to improve the evaluation.
To evaluate a predicted program, we execute it on the test inputs and compare the outputs with the ground truth.

However, checking the exact equivalence of outputs can inadvertently reject correct programs. 
Many problems involve floating point arithmetic, and many return values are acceptable since they are close to the ground truth answer, but they are not exactly equal.
Some problems require random outputs, e.g., generating 100 samples from a distribution, and even executing the reference solution twice can lead to different results.
Many problems do not fully specify all the parameters, e.g., the color scheme for the output figure in the \code{\mpl} library, or the hyper-parameters of a learning algorithm in \code{\sk}; therefore, programs with different parameters can satisfy the requirement, returning values that are different.
In all these cases,
we relied on the best judgment of our expert annotators to implement the metric for each problem, which sometimes involves complicated techniques, such as using statistical tests to handle randomness.
See more examples in Appendix~\ref{sec:examples-appendix}.

\paragraph{Surface-Form Constraints.}
Functional correctness alone is insufficient.
For example, vectorized operations can be expanded using for-loops, which, however, are inefficient and do not meet the requirement of the problem.
Therefore, we introduced additional surface-form constraints that require the presence/absence of specific APIs for keywords.
Notably, such a check is different from the standard surface-form metrics such as CodeBLEU~\citep{codebleu}, which requires the whole model prediction to be uniformly similar to a reference solution; 
instead, \dscg precisely targets small but important parts of surface form.

\subsection{Perturbation to Defend Against Memorization} \label{sec:perturb}

Many models are pre-trained on web text and hence memorize its content \citep{data_leak_memo, extract_train};
therefore, they might answer our problems correctly by simply recalling the solutions seen during pre-training if they were trained on \SO or derivative sites.
We demonstrate this effect on \npHundred,~\footnote{\url{https://github.com/rougier/numpy-100}} a problem set of 100 \code{\np} problems with solutions that are copied several thousand times on \github.
When prompted to answer a selected subset of 20 problems, \codexTwo achieves 72.5\% pass@1 accuracy.\footnote{The fraction of \codexTwo samples that are correct.}

However, if the model truly knows how to solve those problems, it should be able to solve similar problems at the same level of difficulty.
This motivates us to perturb the problems in two ways: surface perturbations and semantic perturbations.
For surface perturbations, we paraphrased the problem or modified the code context in the problem, but the reference solution should stay the same after the perturbation;
for example, changing from ``Create a 5x5 matrix \dots'' to ``I need a matrix sized 5x5 \dots''.
For semantic perturbations, we changed the semantics of the reference solution without changing its difficulty ;
for example, asking for ``\textit{min}'' instead of ``\textit{max}'' in the problem.
We provide more detailed categories in \Cref{tab:example of perturbation}.
In all of these cases, the difficulty of the problem does not change for humans.

\begin{table}[h]
    \centering

\begin{tabular}{@{}c|cc|c@{}}
\toprule
Origin    & Surface     & Semantic     & Avg. Perturbation \\ \midrule
72.5   & 50.8   & 23.6   & 40.6 \\
\bottomrule
\end{tabular}
\caption{The performance of \codexTwo on \npHundred.}
\label{tab:numpy_100}
\vspace{-5pt}
\end{table}

\begin{table*}[h]
\centering
\renewcommand\arraystretch{1.1}
 \resizebox{\linewidth}{!}{
\begin{tabular}{l|ccccccc|c}

\toprule
                       & \pd  & \np  & \mpl  & \sk  & \scp & \tf & \py  & Total/Avg.    \\ \hline
Problem               & 291  & 220  & 155   & 115  & 106  & 45  & 68  & 1000         \\
\hspace{10pt} Origin                         & 100  & 97   & 111  & 46  & 58   & 17   & 22    & 451          \\
\hspace{10pt} Surface Perturbation           & 24   & 22   & 0    & 57  & 11   & 11   & 27    & 152          \\
\hspace{10pt} Semantic Perturbation          & 88   & 51   & 44   & 9   & 20   & 12   & 11    & 235          \\
\hspace{10pt} Difficult Rewrite              & 79   & 50   & 0    & 3   & 17   & 5    & 8     & 162          \\ \midrule
% Execution-based Test   & 156  & 261  & 58   & 36   & 91   & 86   & 155  & 843          \\
\% Surface-Form Constraints    & 12.0   & 36.4   & 0     & 27.8   & 17.9   & 20.0   & 27.9   & 19.4          \\
Avg. Test Cases        & 1.7   & 2.0   & 1.0   & 1.5   & 1.6   & 1.6   & 1.7   & 1.6         \\
% Keywords                & 1.02 & 1.13 & 1.00 & 1.50 & 2.53 & 1.07 & 0 & 1.22         \\ 
\midrule

Avg. Problem Words   & 184.8 & 137.5 & 21.1 & 147.3 & 192.4 & 133.3 & 133.4 & 140.0 \\
% Avg. Code Context Words & 82.0  & 48.1  & - & 51.4  & 67.4  & 54.6  & 33.7  & - \\
Avg. Lines of Code Context & 9.0 & 8.3 & 6.9 & 11.0 & 10.2 & 9.2 & 9.0 & 8.9\\
% Avg. Lines of NL Context & 20.37 & 10.16 & - & 11.36 & 12.25 & 10.96 & \phantom{0}9.22 & -\\
Avg. Lines of Code Solution  & 5.4 & 2.5 & 3.0 & \phantom{0}3.3 & \phantom{0}3.1 & 4.1 & 2.1   & 3.6             \\

\bottomrule
\end{tabular}
}
\caption{
Detailed statistics of \dscg. 
}
\label{tab:statics}
\end{table*}

We manually applied these perturbations to \npHundred and show the result on \Cref{tab:numpy_100}. 
Although the difficulty level remains the same to human users, the performance of \codexTwo drops to 40.6\% after perturbation (50.8\% on surface perturbations and 23.6\% on semantic perturbations).
Furthermore, in 36\% of the cases, the model still predicted the original answer of the problem after the semantic perturbation, implying that the model is solving the original problems by memorizing their corresponding solutions.
Therefore, we could significantly overestimate model performance if we test them on problems directly taken from the web. (See \Cref{sec:np100-appendix} for more details)

Therefore, to proactively prevent memorization, we applied the above two perturbations to \dscg problems.
Perturbation is a labor-intensive process.
Even for a simple perturbation from \emph{min} to \emph{max}, our annotators needed to edit all mentions of \emph{min}, \emph{smallest}, \emph{minimum} to make the problem coherent, and updated the code context, reference solution, and our evaluation metric accordingly.

Finally, to make \dscg more challenging, we additionally introduced several semantic perturbations that increase the difficulty on purpose (``Difficult Rewrite'' in~\Cref{tab:example of perturbation}).

\subsection{Quality Assurance} \label{sec:qa}

To ensure the quality of our benchmark, each problem, reference solution, and automatic multi-criteria evaluation were reviewed by at least three expert annotators familiar with the library.
Additionally, we ``red teamed'' our automatic evaluation by requiring it to reject all programs known to be incorrect, e.g., solutions to semantically perturbed problems (see \Cref{fig:pipeline}).
After the quality review, we also quantitatively measured the evaluation quality by examining whether our multi-criteria automatic metric can reject incorrect Codex-002 predictions (more details in \Cref{sec:all-stats}).

\section{Dataset Statistics}
\label{sec:all-stats}

We provide detailed dataset statistics in \Cref{tab:statics}.
\dscg contains 1000 problems originating from 451 unique \SO problems. 
To defend against potential memorization, more than half of the \dscg problems are modified from the original \SO problems (\Cref{sec:perturb}); they include 152 surface perturbations, 235 semantic perturbations, and 162 difficult rewrites.

\dscg has carefully designed testing methods, checking both execution semantics and surface-form constraints.
For each problem, there are 1.6 test cases (manually annotated corner test cases) on average, and 19.4\% of them are accompanied by surface-form constraints.
The average of problem words in \dscg is 140.
On average, the reference solution contains 3.6 lines of code.
\Cref{tab:statics} shows the library breakdown statistics: 
Most libraries have a similar distribution except \code{\mpl} because we adopted a different problem format due to its multimodal nature.

\Cref{tab:comparison} compares \dscg to other datasets. 
Notably, the average number of words per problem in \dscg is much larger than other data science related datasets (e.g., DSP,~\citealt{JuPyT5-DSP} and CoNaLa,~\citealt{yin2018mining}).
More importantly, the problems in \dscg represent more diverse and naturalistic intent and context formats that cannot be seen in any other datasets.
Unlike generic Python code generation benchmarks (MBPP,~\citealt{austin2021program} and HumanEval,~\citealt{Codex-HumanEval}), we note that data science code generation benchmarks have fewer test cases since the annotators need to define program inputs with complex objects such as square matrices, classifiers, or dataframes rather than simple primitives, such as floats or lists.
Nevertheless, as we will show next, even a few test cases suffice for \dscg.

\begin{table*}[t]
    \centering
    \renewcommand\arraystretch{1.2}
     \resizebox{\linewidth}{!}{
\begin{threeparttable}
\begin{tabular}{@{}lccccccl@{}}
\toprule
Dataset   & Problems & Evaluation    & Avg. Test Cases    & Avg. P Words  & Avg. Lines of Code Solution    & Data Source \\
\midrule
HumanEval & 164       & Test Cases    & \phantom{0}7.7     & \phantom{00}23.0  & \phantom{0}6.3    & Hand-Written \\
MBPP      & 974       & Test Cases     & \phantom{0}3.0    & \phantom{00}15.7  & \phantom{0}6.7    &  Hand-Written \\ 
APPS      & 10000     & Test Cases     & 13.2              & \phantom{0}293.2  &           18.0    & Competitions \\

\midrule
JuICe     & 1981      & Exact Match + BLEU          & -                  & \phantom{00}57.2  & \phantom{0}3.3    & Notebooks \\
DSP       & 1119      & Test Cases     & \phantom{0}2.1    & \phantom{00}71.9  & \phantom{0}4.5    & Notebooks \\
CoNaLa    & 2879       & BLEU          & -                  & \phantom{00}13.8  & \phantom{0}1.1    & \SO 
\\
\midrule
\dscg     & 1000      & \shortstack[c]{Test Cases + \\Surface-Form Constraints}  & \phantom{0}1.6    & \phantom{0}140.0    & \phantom{0}3.6 & \SO \\
\bottomrule
\end{tabular}
\end{threeparttable}
}
\caption{
Comparison of \dscg to other benchmarks. 
The first three benchmarks target general Python usage and the next three involve data science code generation.
\dscg adapts realistic problems from \SO and checks both execution semantics and surface-form constraints.
}
\label{tab:comparison}
\end{table*}

We evaluate our multi-criteria automatic metric by checking whether it can reject incorrect solutions.
We randomly sampled 10 problems from each library and sampled 40 predictions from \codexTwo for each problem (2800 problem-code examples in total).\footnote{We use a higher temperature of 0.7 compared with 0.2 in Section \ref{sec:exp_setup} to get more diverse predictions.}
We run our automatic metric on all the sample predictions, review the predictions manually, calculate how often they disagree, and report the following four quantities:

\begin{itemize}
    \item Sample Level False Discovery Rate: among all predicted samples that \textbf{pass} our automatic evaluation, 1.8\% of them are \textbf{incorrect} according to our annotator.
    \item Sample Level False Omission Rate: among all predicted samples that \textbf{do not pass} our automatic evaluation, 0.5\% of them are \textbf{correct} according to our annotator.
    \item Problem Level False Positive Percentage: among all problems, 5.7\% of the problems contain at least one incorrect sample prediction that pass our automatic metric.
    \item Problem Level False Negative Percentage: among all problems, 5.7\% (it happens to be the same as the above) problems contain at least one correct sample prediction that fails to pass our automatic metric.
\end{itemize}

Generally, problem-level measures are especially stringent since they require correctly judging all predictions among the 40 sample predictions. 
While an apple-to-apple comparison with other datasets is not possible due to the difference in the underlying model and benchmark construction method (as a point of reference, \citet{alphacode} find the problem Level False Positive Percentage to be 60\% on APPS~\citep{hendrycksapps2021}),
these measures reflect that \dscg is reliable.\footnote{Some problems in APPS might apply quite similar tests, and some problems may have even as few as 2 or 3 test cases in the test split. Thus, insufficient test coverage probably happens though there are more test cases in average~\citep{alphacode}. }

\section{Benchmarking State-of-the-Art Models}\label{sec:experiments}

We used \dscg to benchmark five pre-trained code models from three different families. 
The best model \codexTwo Insertion achieves 43.3\% accuracy, indicating room for improvement. 
We also show the results on the perturbed and unperturbed examples in \Cref{sec:analysis-perturbation}.

\subsection{Prompt Format}\label{sec:prompt-format}
We provide an official prompt format in \dscg because it significantly impacts the performance of pre-trained language models~\citep{zhao2021calibrate}.
Figure~\ref{fig:example-datapoint} shows an example: each prompt starts with a natural language description and then provides a code context; the code context uses HTML-like markers to indicate the location of missing code that a model needs to fill in and provides both left and the right context to the missing code pieces.

We decide to use infilling as our official format because the right context is important to specify the behavior of the program predictions (e.g., the variable name for the result).
More broadly, given that 1) infilling is an important functionality for real-world programming and 2) there is a growing trend in pre-training with the right context~\citep{aghajanyan2022cm3, Incoder, bavarian2022efficient, tay2022unifying}, we expect more future pre-trained models to perform infilling.

On the other hand,  given that many current language models trained on code are not yet capable of infilling, we also provide an official prompt that transfers the right context information into the left context (Figure~\ref{fig:example-completion} and \ref{fig:example-complexcompletion}).
Nevertheless, despite our best effort to design the prompts for left-to-right models, they still lag behind models with infilling capabilities (Section \ref{sec: main_result}).
We conjecture that infilling models are inherently more effective at utilizing the right context information.
Finally, we only have Completion format for \code{\mpl} problems because \code{\mpl} provides global access to the current figure so the right context is not necessary.

\begin{table*}[t]
\centering
\renewcommand\arraystretch{0.95}
\tabcolsep=0.08cm

\resizebox{\linewidth}{!}{
        
\begin{threeparttable}
\tiny
\begin{tabular}{c|c|ccccccc|c}
\toprule
Format & Model         & \pd       & \np       & \mpl      & \sk       & \scp      & \tf       & \py     & Overall    \\ \midrule
{\multirow{4}{*}{\shortstack{Left-to-right\\ Completion}}} 
        & \codexTwo     & 26.5     & 43.1     & 57.0    & 44.8     & 31.8     & 39.3     & 41.8   & 39.2   \\
{}      & \codexOne     & \phantom{0}9.4      & 26.6     & 41.8     & 18.5     & 15.0     & 17.2     & \phantom{0}9.7   & 20.2   \\
{}      & \codexCush    & \phantom{0}7.9      & 21.8     & 40.7     & 18.0     & 11.3      & 12.2      & 12.4   & 18.1   \\
{}      & \codegen-6B   & \phantom{0}1.9      & 12.1      & 18.6     & \phantom{0}5.8     & \phantom{0}7.4      & 12.8      & \phantom{0}3.4    & \phantom{0}8.4   \\ 
{}      & \incoder-6B   & \phantom{0}3.1      & \phantom{0}4.4      &  28.3      & \phantom{0}2.8     & \phantom{0}2.8      & \phantom{0}3.8      & \phantom{0}4.4    & \phantom{0}7.4   \\ 
\midrule
{\multirow{2}{*}{Insertion}} 
        & \codexTwo     & 30.1     & 46.5     & 57.0\tnote{*}  & 53.7     & 34.8     & 53.4     & 47.7    & 43.3   \\
{}      & \incoder-6B   & \phantom{0}2.9      & \phantom{0}4.6      & 28.3\tnote{*}    & \phantom{0}3.1      & \phantom{0}3.1      & \phantom{0}7.8      & \phantom{0}3.2    & \phantom{0}7.5    \\
\bottomrule
\end{tabular}
        
\end{threeparttable}
}
\caption{pass@1 accuracy with $40$ samples generated for each problem. The upper part shows accuracy on the left-to-right Completion format, while the lower part shows the results of Insertion format. 
The rightmost ``Overall'' columns show the average accuracy on 1000 problems from all libraries. 
\dscg is able to differentiate the capabilities of different models and there is substantial room for improvement even for the best \codexTwo model.
${}^{*}$: \code{\mpl} problems do not have the right context so Completion and Insertion formats are the same.
}
\label{tab:main-results}
\end{table*}

From now on, we refer to the infilling prompt format as Insertion format and the left-context-only format as Completion format.

\subsection{Experimental Setup} \label{sec:exp_setup}

\paragraph{Models.} We experiment with three families of pre-trained models: Codex, \incoder~\citep{Incoder}, and \codegen~\citep{codegen}.
For the Codex models, we experiment with codex-davinci-002 (\codexTwo), codex-davinci-001 (\codexOne), and codex-cushman-001 (\codexCush).
For \incoder and \codegen, we experiment with the $6$B parameters models.
Among these models, Codex and \codegen models are trained to predict the right context while \incoder models are trained for both left-to-right generation and infilling.
In addition, \codexTwo also supports infilling, although the exact model training details are not disclosed.

\paragraph{Implementation Details.} We generate $40$ samples for each \dscg problem with temperature set to $0.2$, top-p cutoff set to  $0.95$, and max generation length set to 1024.
We set the stop sequence tokens to ``</code>'' and ``\# SOLUTION END''. 
These samples are used in the unbiased estimator of pass@1.
For \dscg, evaluating generated codes does not require special computational resources like GPUs.

\subsection{Main Results} \label{sec: main_result}
\Cref{tab:main-results} displays the pass@1 accuracy on \dscg.
We find that \dscg can differentiate models with different capabilities.
The best model \codexTwo achieves a nontrivial but far-from-perfect average accuracy of 43.3\%, indicating substantial room for improvement.
In contrast, other models like \codegen-6B or \incoder-6B have much worse overall performance, with accuracy lower than 5\% on some libraries.
Qualitatively, these smaller models often cannot correctly follow the prompt instruction, generating additional comments instead of the required code.
Future ablation is needed to understand the underlying cause for this performance gap, which could be the difference in model size, lack of instruction tuning, or the difference in pre-training data.

In addition, we observe that model accuracy varies across different libraries.
This speaks to the importance of a holistic evaluation of multiple data science libraries because performance in a specific library may not directly generalize to other libraries.

Moreover, we find that Insertion format often leads to better performance.
The same \codexTwo model has a 4.1\% average accuracy improvement when used with Insertion format than used with Completion format.
This shows the importance of the infilling capability for data science code completion.

\subsection{Results by Perturbation}
\label{sec:analysis-perturbation}

In \Cref{sec:perturb}, we demonstrated the risk of memorizing the solutions on the \npHundred problem set; 
do we observe the same effect on \dscg?
To investigate this, we applied surface perturbations (i.e., the problem changes but the reference solution does not change) and semantic perturbations (the reference solution will change) to the problems in \dscg.

\begin{table*}[t]
\centering
\renewcommand\arraystretch{1.1}

\resizebox{\linewidth}{!}{
\begin{threeparttable}
\begin{tabular}{l|llllll|l}
\toprule
                    & \pd   & \hspace{-6.5pt}\np     & \hspace{-14pt}\sk   & \hspace{-2.8pt}\scp     & \hspace{-14.5pt}\tf    & \hspace{-8pt}\py    & \hspace{0pt}Overall               \\
\midrule
% Total               & \hspace{5.5pt}22.7     & 40.4     & 39.6     & 28.8     & 35.9        & 38.5                        \\ \hline
Origin\textsubscript{surface}  & \hspace{5.5pt}37.3     & 61.2     & 52.6     & 33.0     & 64.9    & 64.8    &   \hspace{6.5pt}53.2          \\

Surface             & \hspace{5.5pt}31.9~{\fontsize{7pt}{0}\selectfont\color{orange} $-5.4$}     
                    & 58.4~{\fontsize{7pt}{0}\selectfont\color{orange} $-2.8$}
                    & 55.7~{\fontsize{7pt}{0}\selectfont\color{orange} $+3.1$}
                    & 32.1~{\fontsize{7pt}{0}\selectfont\color{orange} $-0.9$}
                    & 58.0~{\fontsize{7pt}{0}\selectfont\color{orange} $-8.9$}
                    & 50.0~{\fontsize{7pt}{0}\selectfont\color{orange} $-14.8$}
                    & \hspace{6.5pt}49.8~{\fontsize{7pt}{0}\selectfont\color{orange} $-3.4$}
                    \\ \hline
% $\Delta_{surface}$  & -6.15     & -4.32     &  \phantom{-}0.61     & -4.32     & -5.45         & -7.87                 \\ \hline

Origin\textsubscript{semantic} & \hspace{5.5pt}36.8     & 56.7     & 60.6\tnote{*}     & 40.3     & 71.3  & 65.1  & \hspace{6.5pt}47.2
\\

Semantic            & \hspace{5.5pt}33.2~{\fontsize{7pt}{0}\selectfont\color{orange} $-3.6$}  
                    & 49.0~{\fontsize{7pt}{0}\selectfont\color{orange} $-7.7$} 
                    & 38.9\tnote{*}~{\fontsize{7pt}{0}\selectfont\color{orange} $-21.7$}
                    & 34.3~{\fontsize{7pt}{0}\selectfont\color{orange} $-6.0$} 
                    & 42.5~{\fontsize{7pt}{0}\selectfont\color{orange} $-25.8$}      
                    & 30.5~{\fontsize{7pt}{0}\selectfont\color{orange} $-34.6$}
                    & \hspace{6.5pt}38.2~{\fontsize{7pt}{0}\selectfont\color{orange} $-9.0$}
                    \\ \hline
% $\Delta_{mini-sem}$ & -3.69     & -6.91     & -15.00    & -7.38     & -25.00        & -27.50                \\ \hline

Origin\textsubscript{difficult}  & \hspace{5.5pt}39.9     & 52.7     & \phantom{0}5.0\tnote{*}     & 58.1     & 73.0\tnote{*}          & 53.8\tnote{*}      & \hspace{6.5pt}46.8          \\

Difficult Rewrite   & \hspace{5.5pt}17.7~{\fontsize{7pt}{0}\selectfont\color{orange} $-22.2$}       
                    & 27.1~{\fontsize{7pt}{0}\selectfont\color{orange} $-25.6$}       
                    & \phantom{0}0.0\tnote{*}~{\fontsize{7pt}{0}\selectfont\color{orange} $-5.0$}
                    & 13.8~{\fontsize{7pt}{0}\selectfont\color{orange} $-44.3$}       
                    & 38.0\tnote{*}~{\fontsize{7pt}{0}\selectfont\color{orange} $-35.0$}
                    & 28.8\tnote{*}~{\fontsize{7pt}{0}\selectfont\color{orange} $-25.0$}
                    & \hspace{6.5pt}21.0~{\fontsize{7pt}{0}\selectfont\color{orange} $-25.8$}
\\
% $\Delta_{big-sem}$  & -19.05    & -28.00    & -9.17     & -34.71    & -52.00        & -34.38
% \\
\bottomrule
\end{tabular}
        
\end{threeparttable}
 }
\caption{
Effect of three different types of problem perturbation. In each subsection, we compare the accuracy of the perturbed problems to that of the original problems.
We observe that although Surface and Semantic perturbations also cause a performance drop on \dscg the performance drop is much smaller compared to that on \npHundred.
${}^{*}$: Numbers are averaged from less than 10 problems.
% \tianyi{This first row "total" does not convey much information and is confusing how it is related to the rest of the table. So removed for now.}
% Fine-grained performance of \codexTwo Insertion on \dscg. \textbf{Note} that we don't apply Surface perturbation to all of the questions. For a fair comparison, Origin\textsubscript{surface} only contains Origins that have Surface perturbations
}
\label{tab:effect of Perturbation}
\vspace{-5pt}
\end{table*}

Table~\ref{tab:effect of Perturbation} shows the results.\footnote{Note that the results are not comparable to \Cref{tab:main-results} since for each kind of perturbation, we only selected a subset of problems to perturb.}
The performance of \codexTwo drops after perturbation (3.4\% on surface perturbations and 9.0\% on semantic perturbations)%\zrq{needs actual number}
~but the drop is much less severe than what we observed on \npHundred.
This indirectly suggests that \codexTwo might have memorized the solution for some \SO problems, but the effect is less severe because they have not been repeated as often as \npHundred on the internet.
Still, we believe problem perturbation to be a useful strategy to defend against memorization by future models proactively.

Additionally, we rewrote some problems to create more \dscg problems by intentionally making them more difficult even for human programmers.
As expected, \codexTwo performs much worse after the rewrite, and we plan to use these problems as a challenge for future models.

We give a preliminary error analysis in \Cref{sec:error}.
\section{Related Work}\label{sec:related}

\paragraph{Natural Language to Code.}
Research on translating natural language to executable forms dates back several decades.
The models have become increasingly capable of producing complex and general programs while requiring fewer human annotations. 
\citet{zelle96geoquery} and \citet{zettlemoyer2007online} translate natural language queries to domain-specific database queries. 
\citet{liang2013dcs} and \citet{berant2013semantic} parse natural language into first-order logic to answer generic knowledge-based questions.
\citet{yu2018spider, scholak-etal-2021-picard} translate natural language problems to general SQL programs and develop models that can generalize across domains. 
While all the works above still need to train their models on the task they evaluate, recently \citet{alphacode, Codex-HumanEval} show that generative models pre-trained on code can produce Python snippets to tackle competitive programming challenges, without any additional human annotations. 
Many other recent works corroborated this finding \citep{codegen, Incoder, xu2022systematic, black-etal-2022-gpt}, and additional techniques at inference time further improve the performance \citep{poesia2022synchromesh, shi2022natural}.

\paragraph{Code Generation Benchmarks.}
As models become increasingly capable, researchers start to build increasingly difficult and general code generation benchmarks.
While \citet{zelle96geoquery} focused only on domain-specific languages, \citet{yu2018spider} builds a Text-to-SQL benchmark that evaluates the capability to write broad-domain SQL programs.
\citet{yin2018mining} evaluates the capability to write short but general Python snippets, while more recent papers \citet{hendrycksapps2021, alphacode} evaluate models' capability to solve competitive programming problems in Python.
If code generation models continue to improve, we expect future researchers to focus on more complex tasks.

At the same time, however, it becomes more difficult to build reliable benchmarks aligned with real-world applications.
Programs are most useful when they are executed; therefore, we need to evaluate their execution semantics, and the best general method so far is still to ask experts to manually write test cases.
Consequently, most benchmarks with test cases focus on competition/interview/ programming challenges \citep{hendrycksapps2021, alphacode}, because these are the only applications where a lot of test cases are already available.
Therefore, most recent papers that evaluate on real-world programs have to rely on unreliable surface-form metrics \citep{codebleu, chen2021plotcoder, xu2022systematic}.
This streetlight effect might incentivize the community to work on problems that are easy to evaluate but not useful in practice. 
In response to this challenge, our paper manually implements a reliable metric for naturally occurring problems.
Future works can consider using models to help humans write useful tests \citep{tufano2020unit}, or formally verify the correctness of a predicted solution \citep{chu2017cosette}.

\section{Conclusion}\label{sec:conclusion}
We propose \dscg, a benchmark for generating code for data analysis. 
Our benchmark 1) contains realistic problems, 2) implements reliable automatic metrics, and 3) proactively defends against memorization strategies. 
We hope \dscg can track the progress of this research area and facilitate fair comparisons between models, and our methods to construct it can inspire other areas where the task is complicated and the ground truth is challenging to evaluate.
\clearpage
\section*{Acknowledgements}
We thank Noah A. Smith, Tianbao Xie, Shuyang Jiang for their helpful feedback on this work.


\bibliography{main}
\bibliographystyle{icml2022}
\appendix
\clearpage
\begin{appendices}

\section{Details on Data Collection}
\label{sec:data-collection-appendix}

\subsection{Problem Selection}
\label{sec:problem-selection-appendix}

\paragraph{Sourcing Popular \SO Problems.}
We leverage \SO to collect representative data science code generation problems on each library. 
To select popular problems, we first removed duplicates and selected problems with at least 1 vote, 1000 views, and accepted answers. 
After this initial filtering, we obtain 15881 \code{\np} problems, 26248 \code{\pd} problems, 1965 \code{\py} problems, 8258 \code{\tf} problems, 4141 \code{\scp} problems, and 4499 \code{\sk} problems.
Next, we performed a stratified sampling on problems from each year to further subsample the problems from \code{\pd} and \code{\tf}.
We designed a threshold for each year's problems differently because older problems naturally have higher votes.
\Cref{tab:parameters-of-problem-selection} displays the criteria we used to filter each year's problem on \code{\pd} and \code{\tf}.

\begin{figure*}[h]
    \centering
    \includegraphics[width=0.9\textwidth]{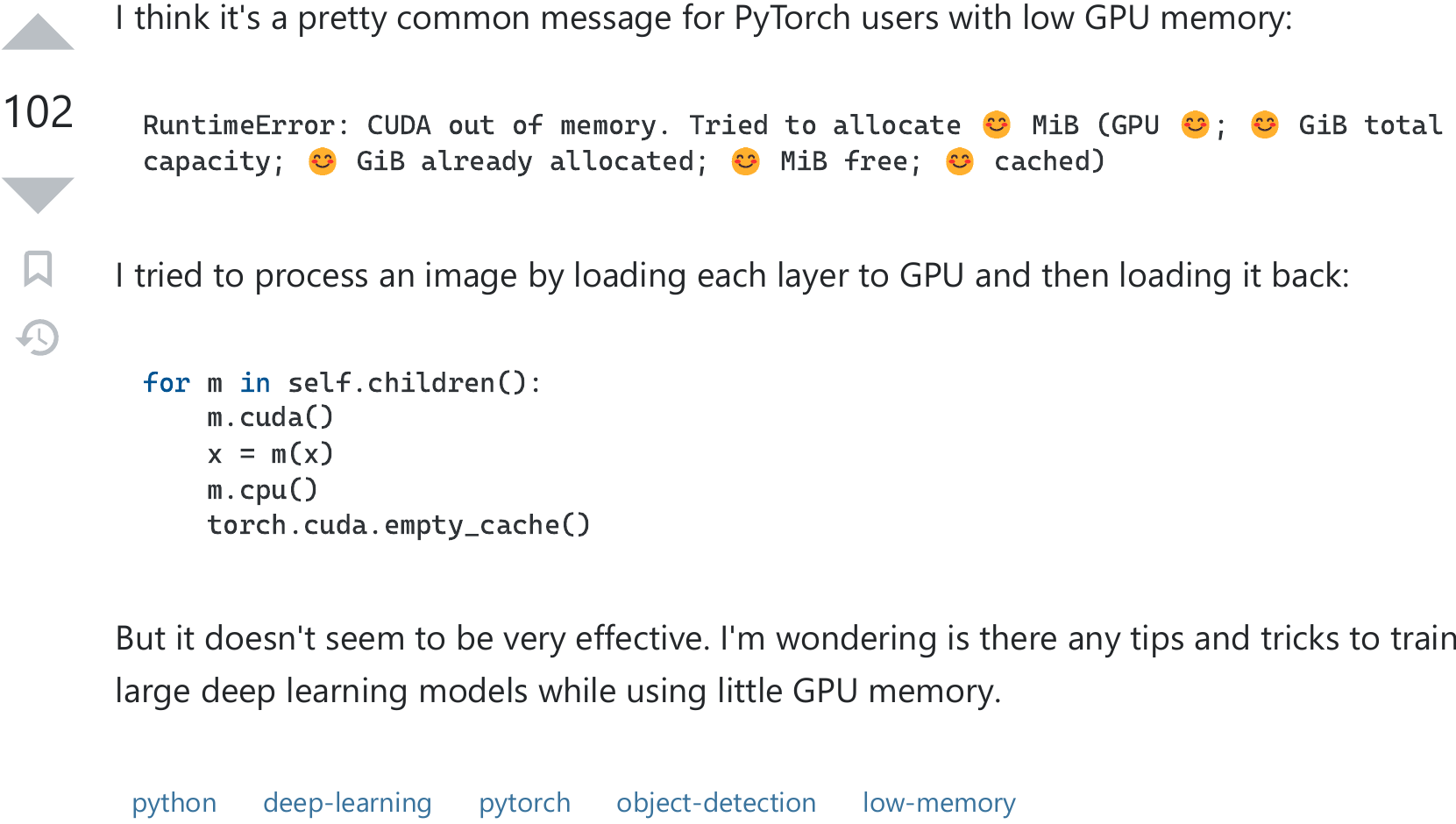}
    \caption{
    An example untestable problem involving hardware problems.
    }
    \label{fig:example-untestable-hardware}
\end{figure*}

\begin{figure*}[h]
    \centering
    \includegraphics[width=0.9\textwidth]{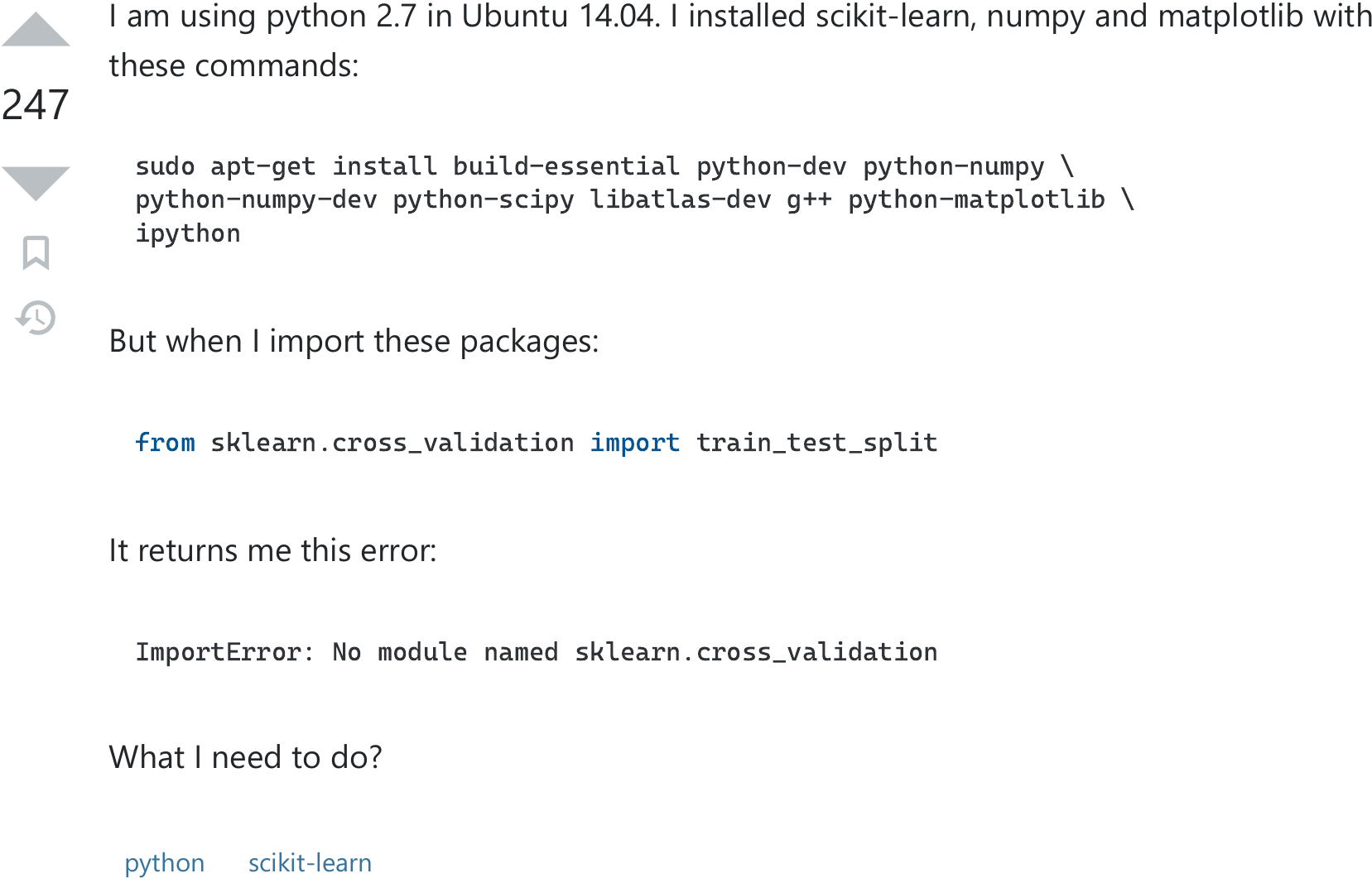}
    \caption{
    An example untestable problem involving software errors.
    }
    \label{fig:example-untestable-software}
\end{figure*}

\begin{figure*}[t]
    \centering
    \includegraphics[width=0.9\textwidth]{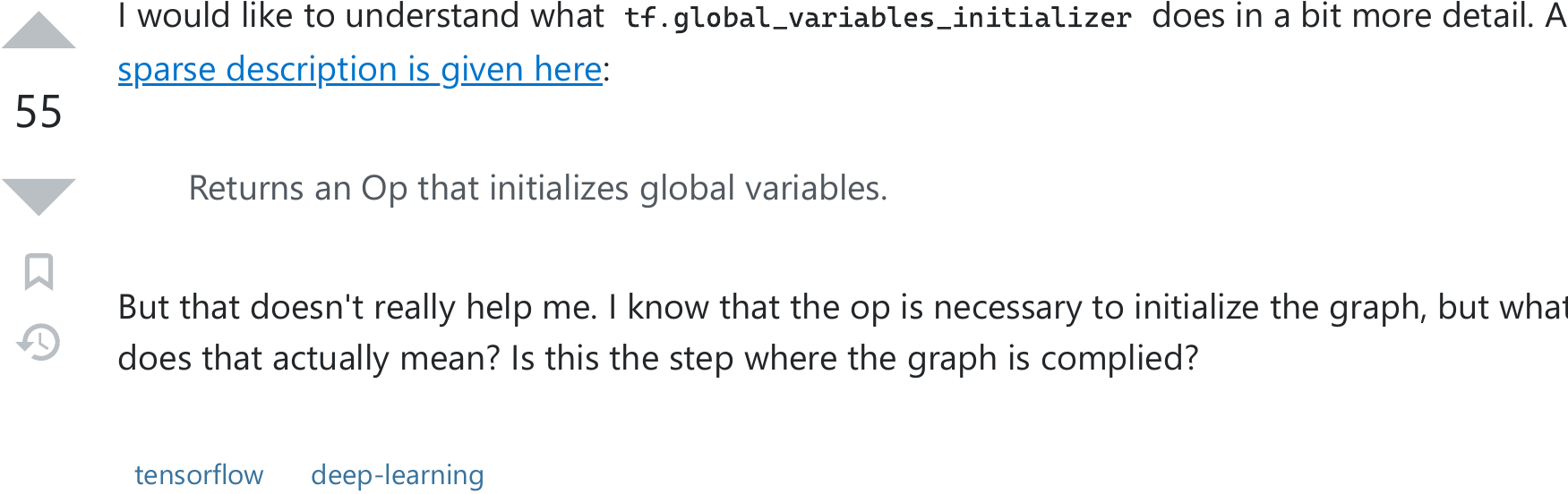}
    \caption{
    An example untestable problem involving explanations.
    }
    \label{fig:example-untestable-explanation}
\end{figure*}

\paragraph{Filtering Suitable Problems.}
From the initial pool of popular problems, our annotators selected problems that are suitable for building \dscg.
Besides the considerations mentioned in Section~\ref{sec:pipeline}, we discuss those problems that are not selected here.
In general, we consider a problem to be unsuitable if our multi-criteria evaluation is not applicable (untestable problems).
For example, we leaved \SO problems involving hardware problems (See Figure~\ref{fig:example-untestable-hardware}), software errors (See Figure~\ref{fig:example-untestable-software}), concrete execution time analysis, etc. out of \dscg.
See Figure~\ref{fig:example-untestable-explanation} for a concrete example where the problem asks for a natural language explanation of a method in \code{\tf}.
We leave incorporating more unsuitable \SO problems for future work.

\paragraph{Controlling Library Version.}
Table \ref{tab:soft_version} details the software versions that we build \dscg with.

\begin{table}[h]
    \centering
\begin{tabular}{@{}cc@{}}
\toprule
Package    & Version\\ \midrule
Seaborn   & 0.11.2 \\
\mpl   & 3.5.2  \\
\np   & 1.21.6  \\
\pd   & 1.3.5  \\
\sk   & 1.0.2  \\
\scp   & 1.7.3  \\
\tf   & 2.10.0  \\
\py   & 1.12.1  \\
\bottomrule
\end{tabular}
\caption{The versions of software in \dscg}
\label{tab:soft_version}
\end{table}

\begin{table*}[t]
\centering
\renewcommand\arraystretch{1.2}
\resizebox{\linewidth}{!}{
\begin{tabular}{llcccccccccccc}
\hline
 & Year & 2011  & 2012  & 2013  & 2014  & 2015  & 2016  & 2017  & 2018  & 2019  & 2020  & 2021  & 2022 \\ \hline
 % \pd  &  c  &  0.2  &  0.2  &  0.25  &  0.25  &  0.25  &  0.45  &  0.45  &  0.45  &  0.65  &  0.65  &  1.0  &  1.0  \\
 \pd &  vote  &  50  &  50  &  14  &  14  &  14  &  4  &  4  &  4  &  2  &  2  &  1  &  1  \\
 &  view  &  5k  &  5k  &  5k  &  5k  & 5k  &  1k  &  1k  &  1k  &  1.1k  &  1.1k  &  1k  &  1k  \\ 
 &  problems  &  2  &  8  &  467  &  494  &  554  &  2139  &  2483  &  1894  &  1985  &  809  &  225  &  8  \\ \hline
 % \tf  &  c  &  -  &  -  &  -  &  -  &  0.4  &  0.45  &  0.5  &  0.6  &  0.7  &  0.8  &  1.0  &  1.0  \\
 \tf &  vote  &  -  &  -  &  -  &  -  &  10  &  5  &  4  &  2  &  2  &  1  &  1  &  1  \\
 &  view  &  -  &  -  &  -  &  -  & 3k  &  2k  &  1k  &  1.6k  &  1.2k  &  1.3k  &  1k  &  1k  \\
 &  problems  &  -  &  -  &  -  &  -  &  100  &  632  &  1136  &  1167  &  1004  &  776  &  185  &  6  \\
\hline
\end{tabular}
}
\caption{The problem selection parameters and the number of result problems of \code{\pd} and \code{\tf}.}
\label{tab:parameters-of-problem-selection}
\end{table*}

\subsection{Example Problems}
\label{sec:examples-appendix}
Here we present an example problem from each of the seven libraries in \dscg to illustrate the challenges we encountered in creating \dscg.

\begin{figure*}[h]
    \centering
    \includegraphics[width=1\textwidth]{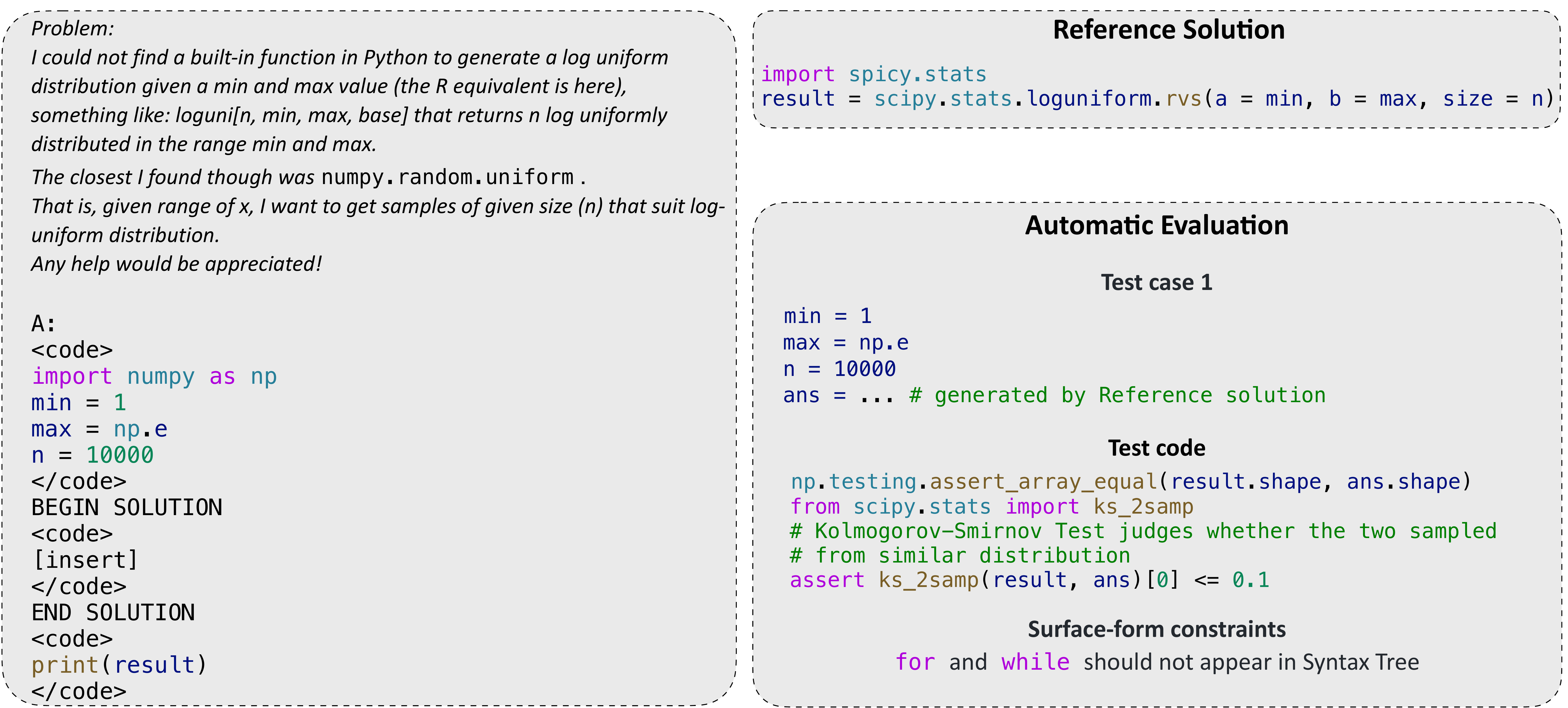}
    \caption{
    \code{\np} example problem involving randomness, requiring the use of a specialist knowledge test.
    }
    \label{fig:example-numpy-251}
\end{figure*}

Figure~\ref{fig:example-numpy-251} shows a \code{\np} problem asking how to generate samples that suit log-uniform distribution. Since the result varies with different solutions and different settings, it's unreasonable to test the equivalence. Instead, we apply the Kolmogorov-Smirnov test that judges whether two groups of samples suit the identical or rather similar population.

\begin{figure*}[h]
    \centering
    \includegraphics[width=0.9\textwidth]{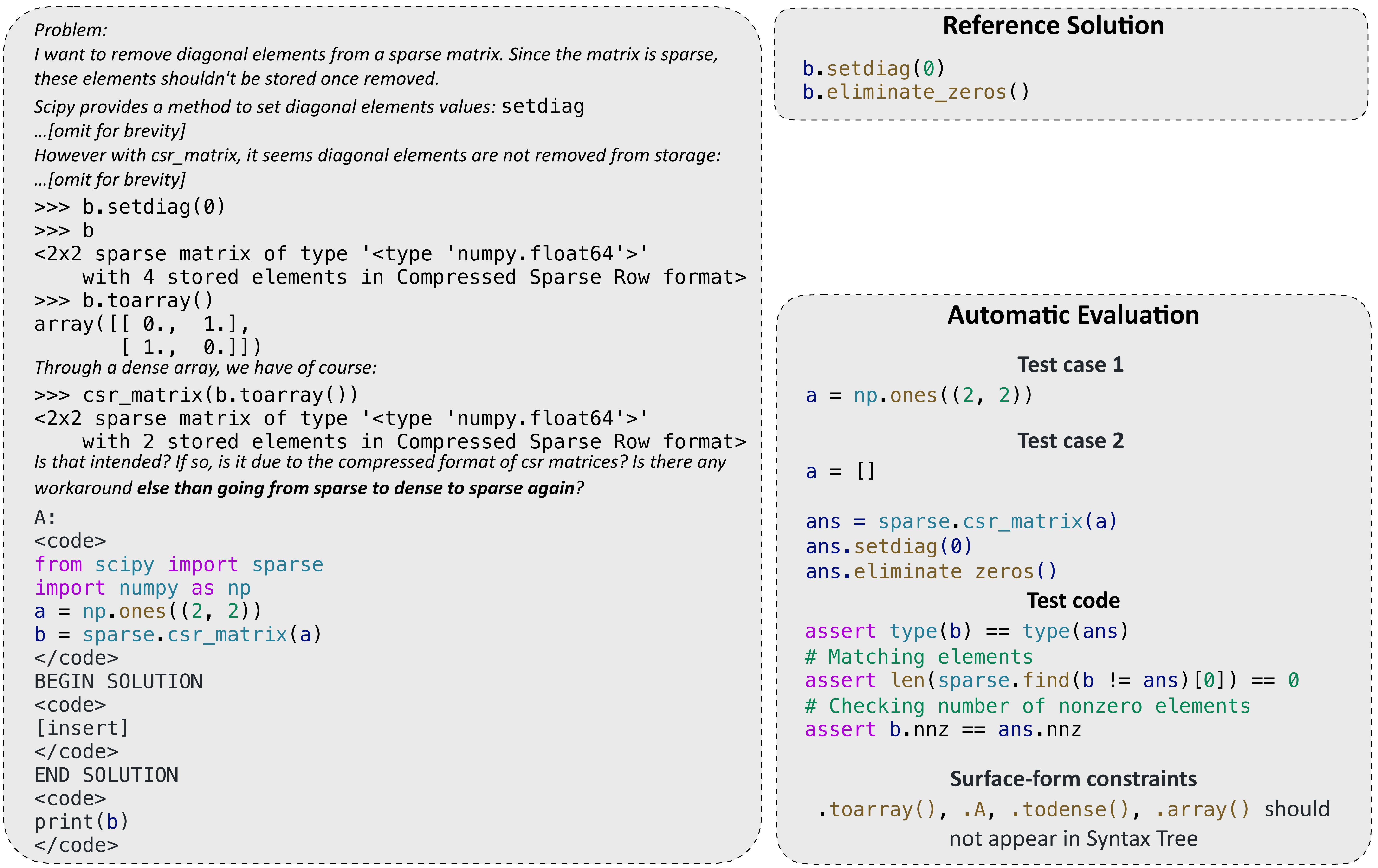}
    \caption{
    An example problem of \code{\scp}. Specific checking on conversion between dense matrix and sparse matrix.
    }
    \label{fig:example-scipy-124}
\end{figure*}

Figure~\ref{fig:example-scipy-124} gives a \code{\scp} problem that describes some trouble with the number of stored elements in a sparse matrix and asks for a solution without repetitive type conversion. Since our self-made assertion that checks the equivalence of two matrices cannot distinguish the difference between stored numbers, we need a special design for this problem. For functional correctness, we check the type of \code{b}, match the elements, and check the number of non-zero elements(\code{nnz}), which is the core of the problem. For surface-form constraints, we reject the use of \code{.toarray()}, \code{.A}, \code{.todense()}, and \code{.array()}, which might attempt to transform a sparse matrix into a dense one.

\begin{figure*}[h]
    \centering
    \includegraphics[width=1\textwidth]{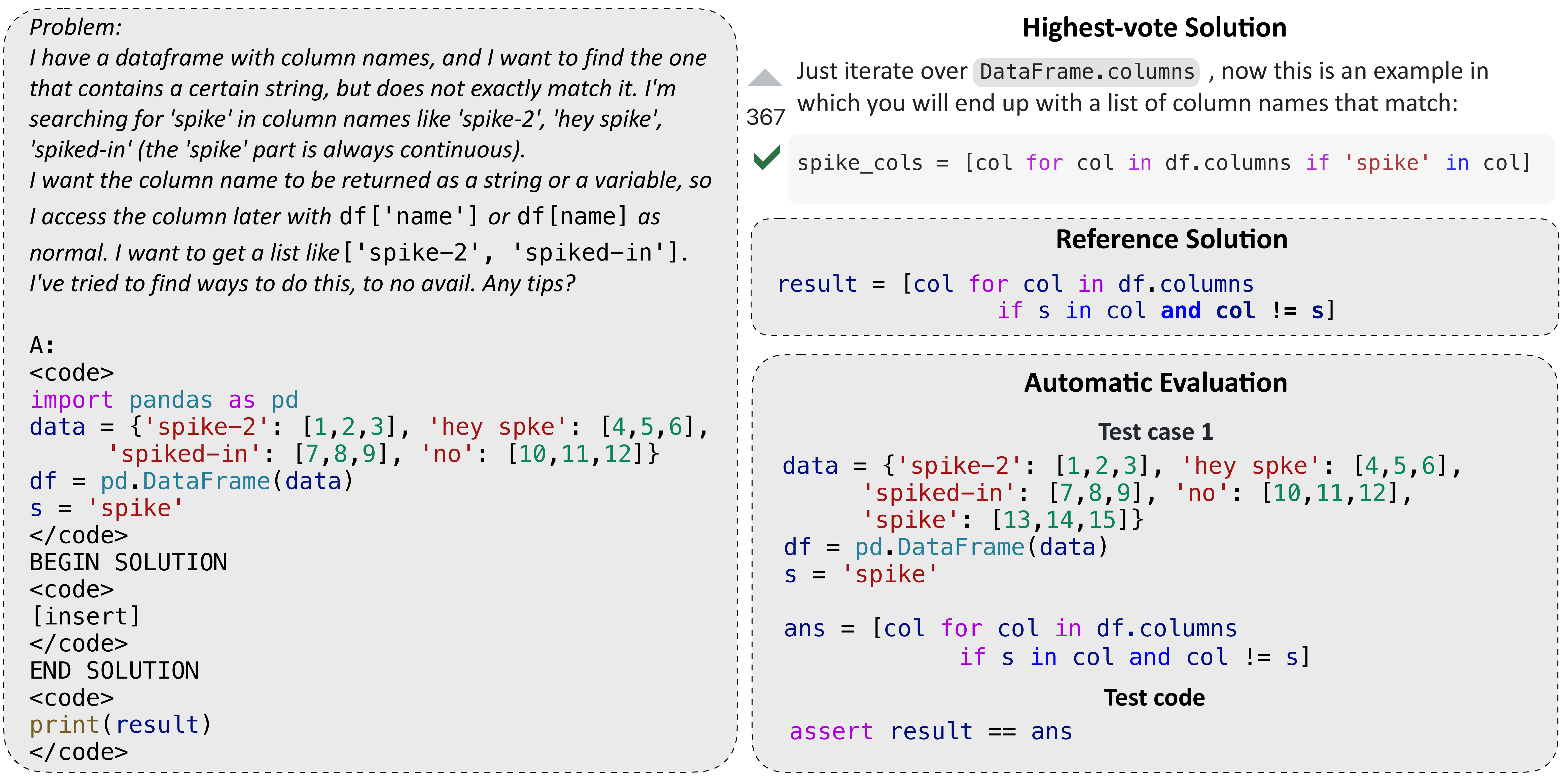}
    \caption{
    An example problem of \code{\pd}. We need to write reference solutions by ourselves because high-vote replies from \SO ignore the requirement ``but does not exactly match it''.
    }
    \label{fig:example-pandas-115}
\end{figure*}

Figure~\ref{fig:example-pandas-115} shows a \code{\pd} problem. We found that the solution with the highest votes ignores the requirement ``but does not exactly match it'' in the description of the problem, and thus we had to fix the bug in our reference solution. Besides, we enhanced the test case to check the point.

\begin{figure*}[htbp]
    \centering
    \includegraphics[width=1\textwidth]{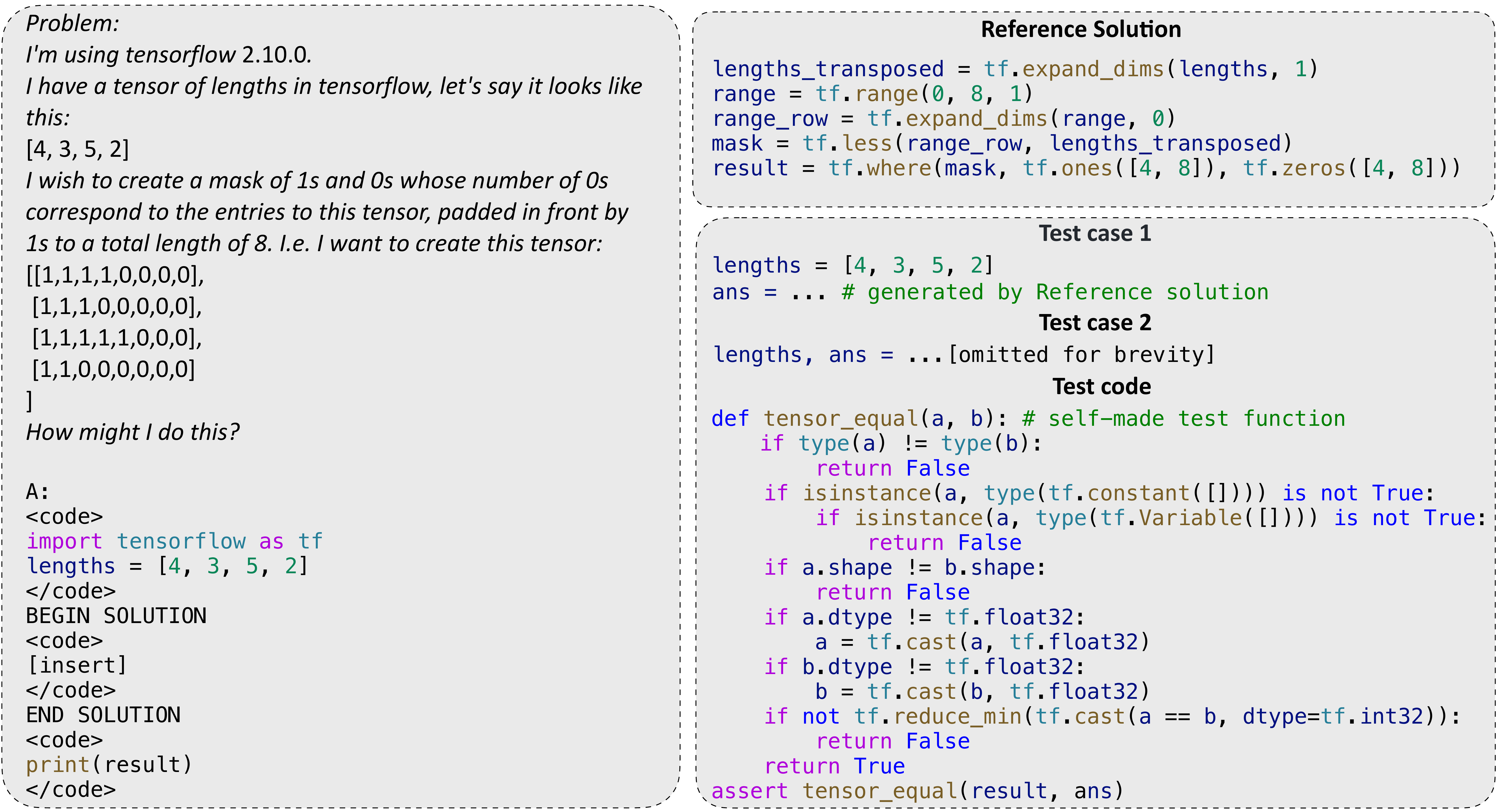}
    \caption{
    An example problem of \code{\tf}. We implemented well-designed test function for tensor comparison.
    }
    \label{fig:example-tensorflow-3}
\end{figure*}

Figure~\ref{fig:example-tensorflow-3} shows a \code{\tf} problem. Since there is no built-in testing function defined in \code{\tf} 2.10.0, we had to design it by ourselves. 

\begin{figure*}[htbp]
    \centering
    \includegraphics[width=1\textwidth]{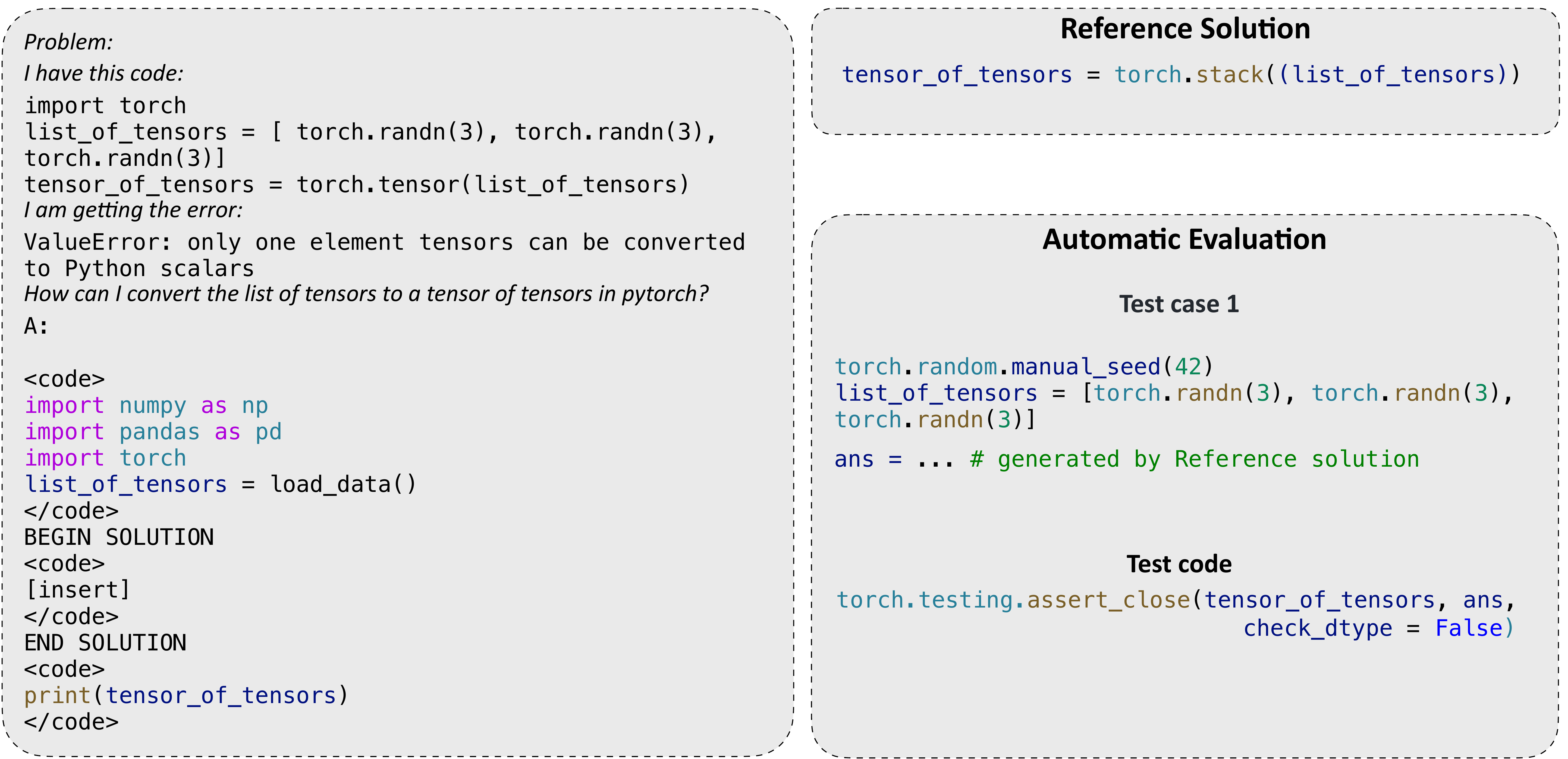}
    \caption{
    An example problem of \code{\py}, with failed attempt and error message given in the description.
    }
    \label{fig:example-pytorch-41}
\end{figure*}

Figure~\ref{fig:example-pytorch-41} demonstrates a \code{\py} problem. Here we use \code{load$\_$data()} to hide the input and let the models learn from the description. The correct solution is not a regular type conversion, as indicated in the error message.

\begin{figure*}[htbp]
    \centering
    \includegraphics[width=1\textwidth]{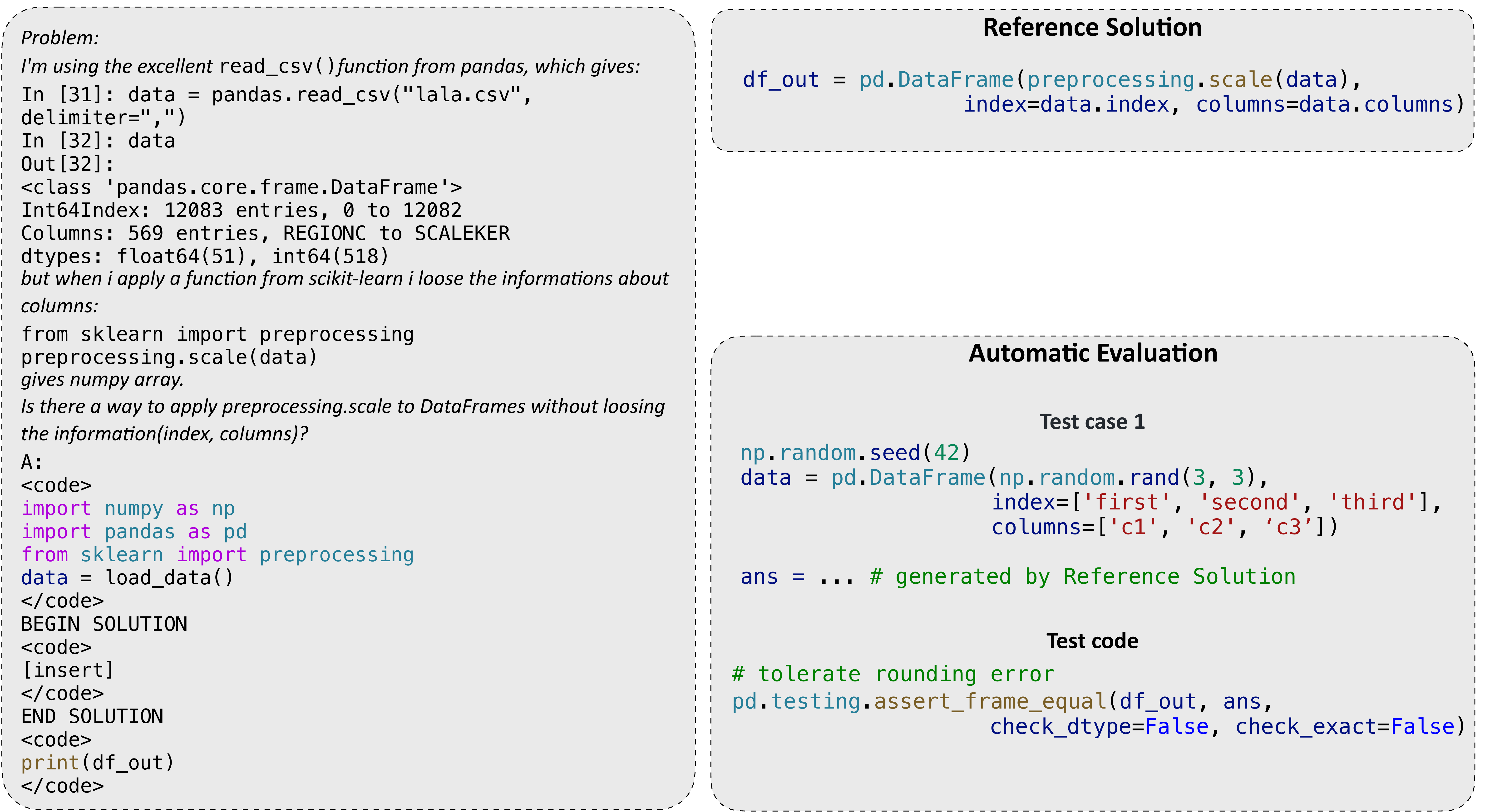}
    \caption{
    An example problem of \code{\sk}, requiring applying sklearn preprocessing method to \pd dataframe.
    }
    \label{fig:example-sklearn-60}
\end{figure*}

Figure~\ref{fig:example-sklearn-60} shows a \code{\sk} problem. It requires applying the preprocessing method defined in \code{\sk} to a \code{\pd} dataframe, and it tests whether the models learn \code{\sk}, \code{\pd}, and their interaction well. Actually, these data science libraries are not independent of others, and this problem exemplifies the interactions.

\begin{figure*}[htbp]
    \centering
    \includegraphics[width=1\textwidth]{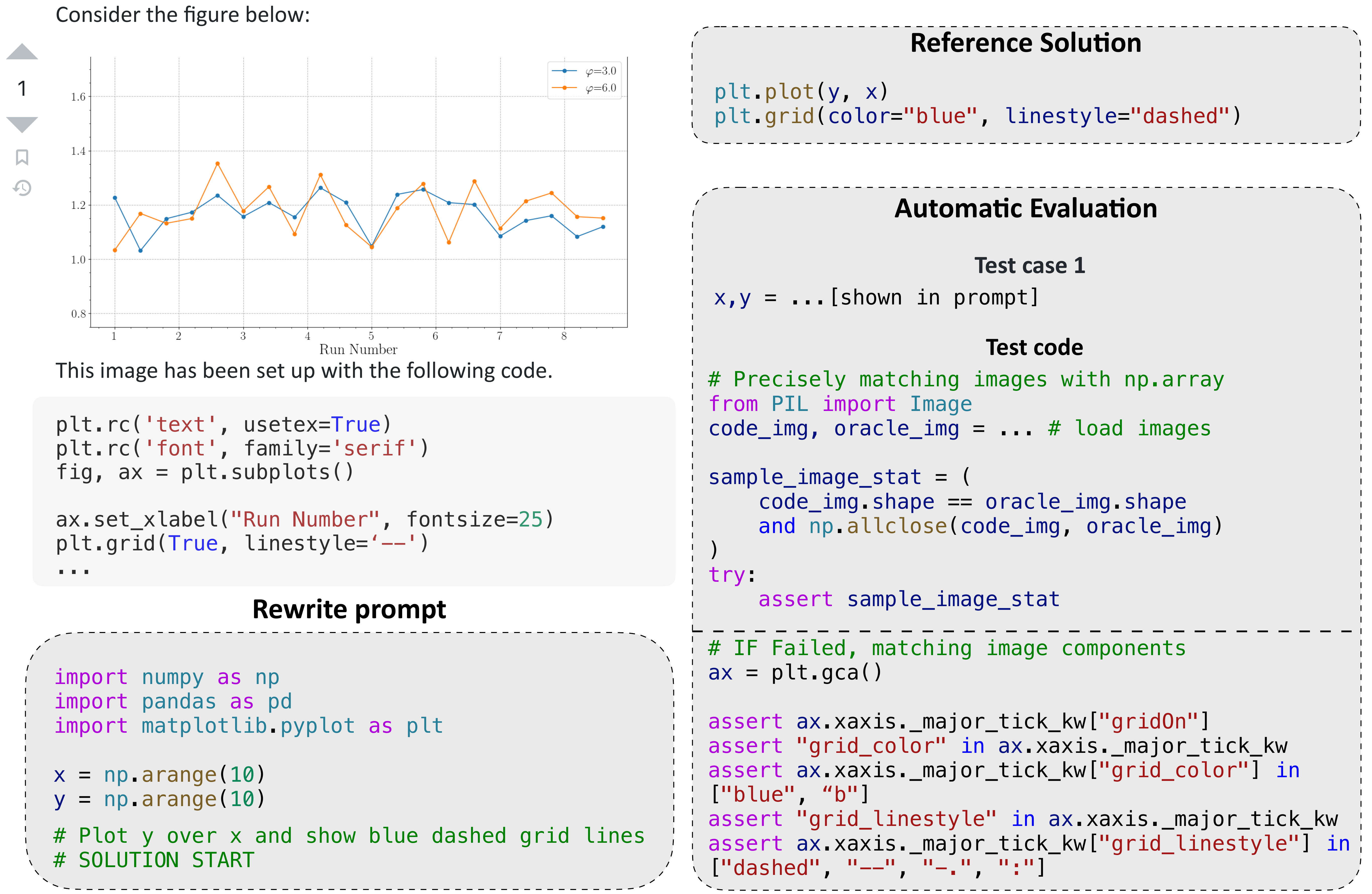}
    \caption{
    An example problem of \code{\mpl}. \code{\mpl} original problems often contain example figures which cannot be processed by current code models. We rewrite original problems into standalone problems in the form of comments.
    }
    \label{fig:example-plotting}
\end{figure*}

Figure~\ref{fig:example-plotting} shows a \code{\mpl} problem. 
Here the original problem on \SO contains an example figure, which cannot be processed by current code models.
We rewrite the original problem into a standalone problem, that is, ``Plot y over x and show blue dashed grid lines''.
The automatic evaluation comes in two parts.
First, it compares the image produced by the generated program with the image produced by the reference program.
If two images match exactly, then the generated program is considered correct.
Otherwise, the automatic evaluation examines the \code{\mpl} axis object and asserts the conditions relevant to the problem specification.
In this example, the assertions are testing the existence of grid lines and the color of the grid lines.

\subsection{Problem Perturbation}
\label{sec:perturbations-appendix}

Here, we give an example for each type of perturbation, as shown in \Cref{tab:example of perturbation}. We highlight the changes we made through perturbations.

\begin{figure*}[h]
    \centering
    \includegraphics[width=0.9\textwidth]{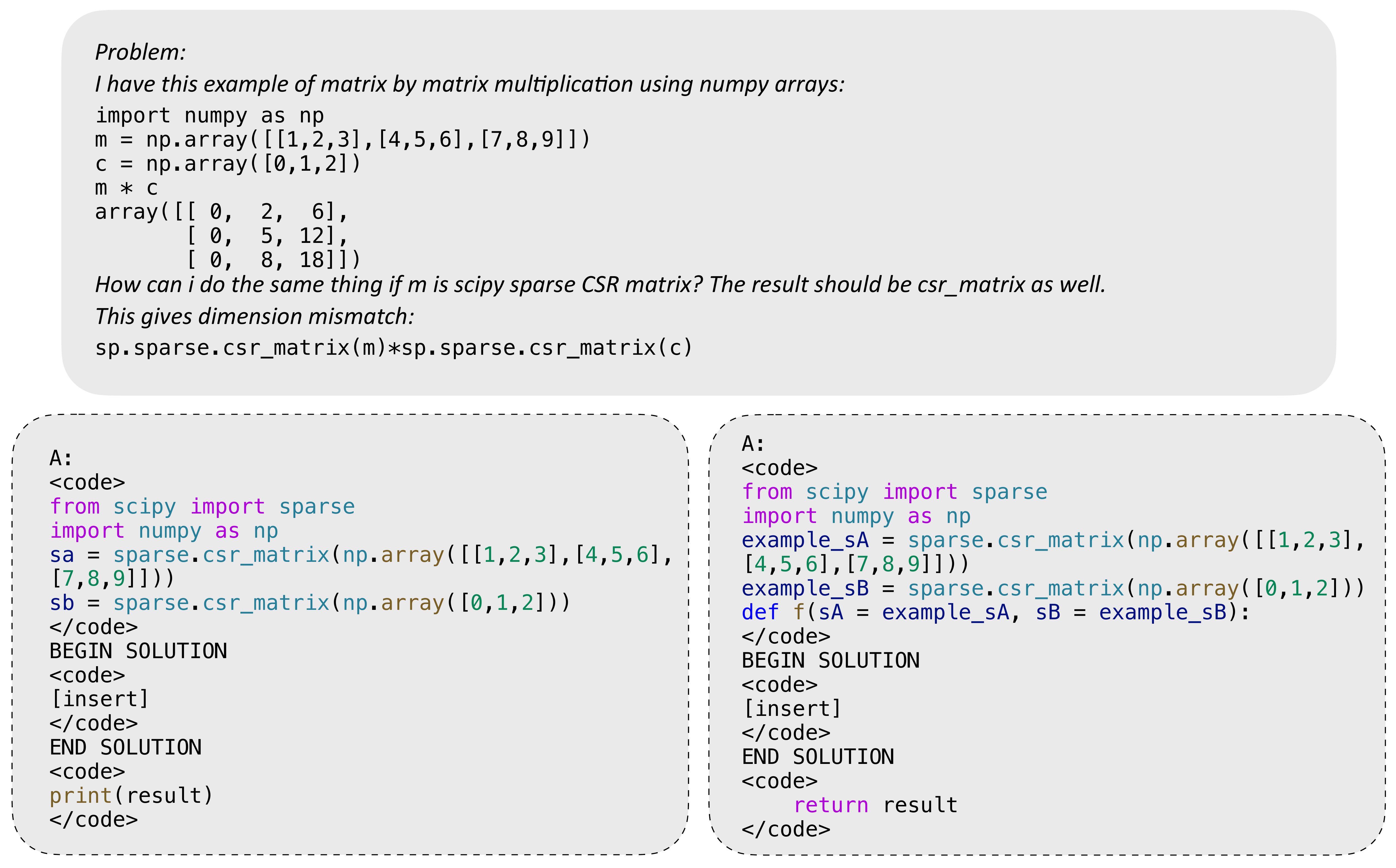}
    \caption{
    An example problem of surface perturbation. We expect model complete the function(on the right).
    }
    \label{fig:example-Function}
\end{figure*}

\begin{figure*}[h]
    \centering
    \includegraphics[width=0.75\textwidth]{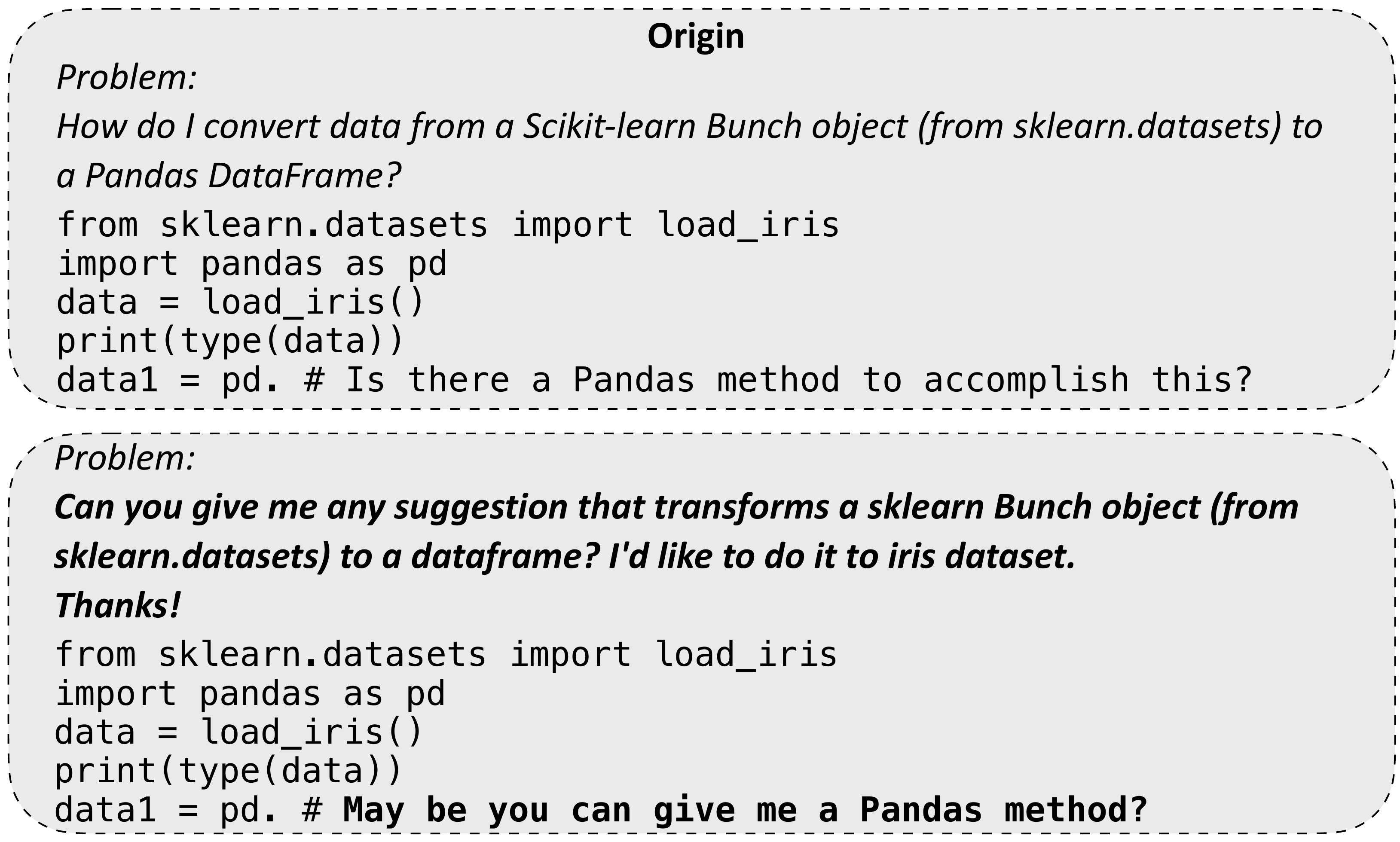}
    \caption{
    An example problem of surface perturbation. The description in the prompt has been paraphrased.
    }
    \label{fig:example-A1}
\end{figure*}

\begin{figure*}[h]
    \centering
    \includegraphics[width=0.75\textwidth]{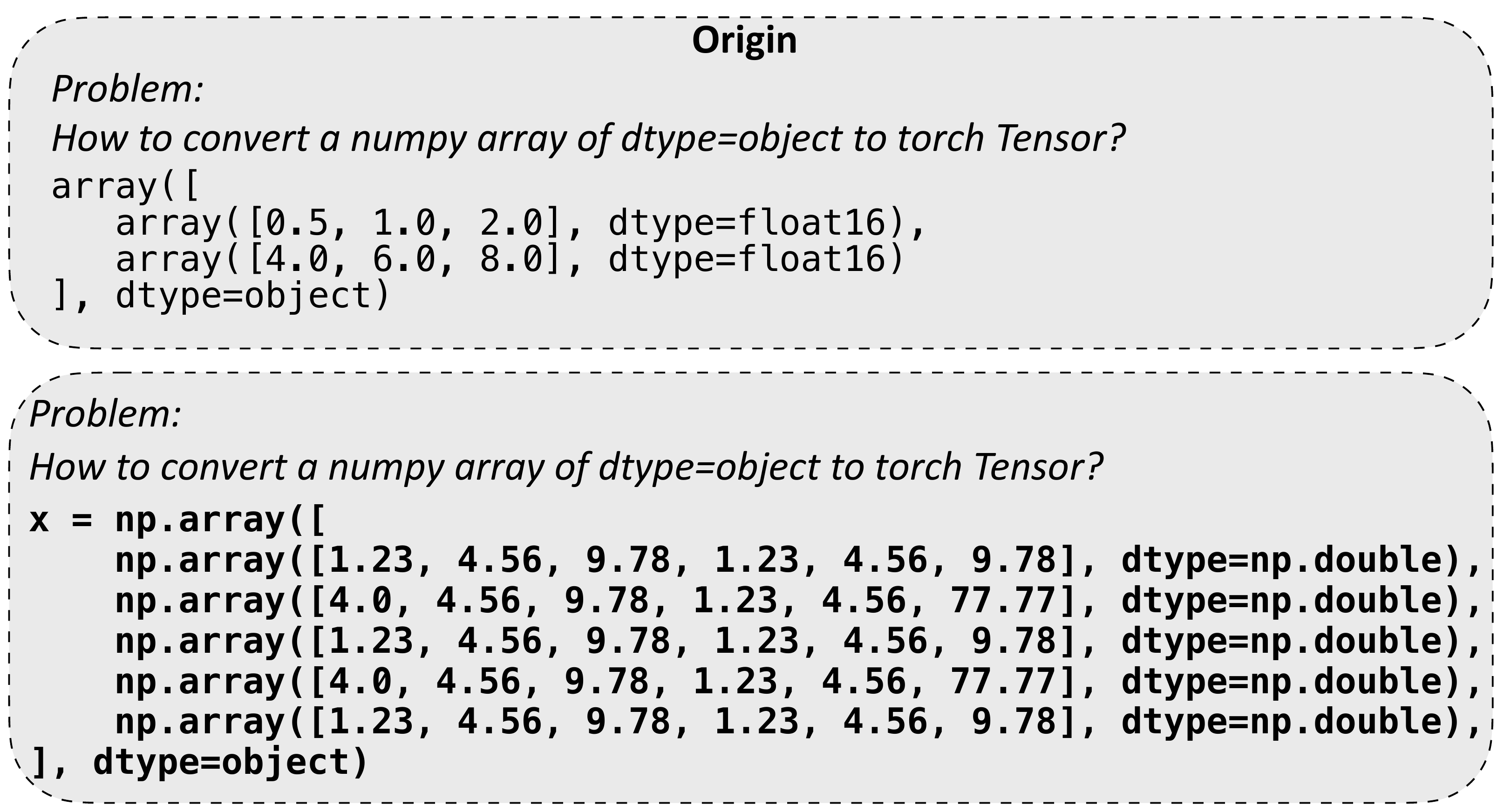}
    \caption{
    An example problem of surface perturbation. The example input in the prompt has been replaced with another one.
    }
    \label{fig:example-A2}
\end{figure*}

Figure~\ref{fig:example-Function}, Figure~\ref{fig:example-A1} and Figure~\ref{fig:example-A2} give examples of surface perturbations, showing code context perturbation,  paraphrasing, and changes in example respectively. The original task hasn't changed.

\begin{figure*}[h]
    \centering
    \includegraphics[width=0.75\textwidth]{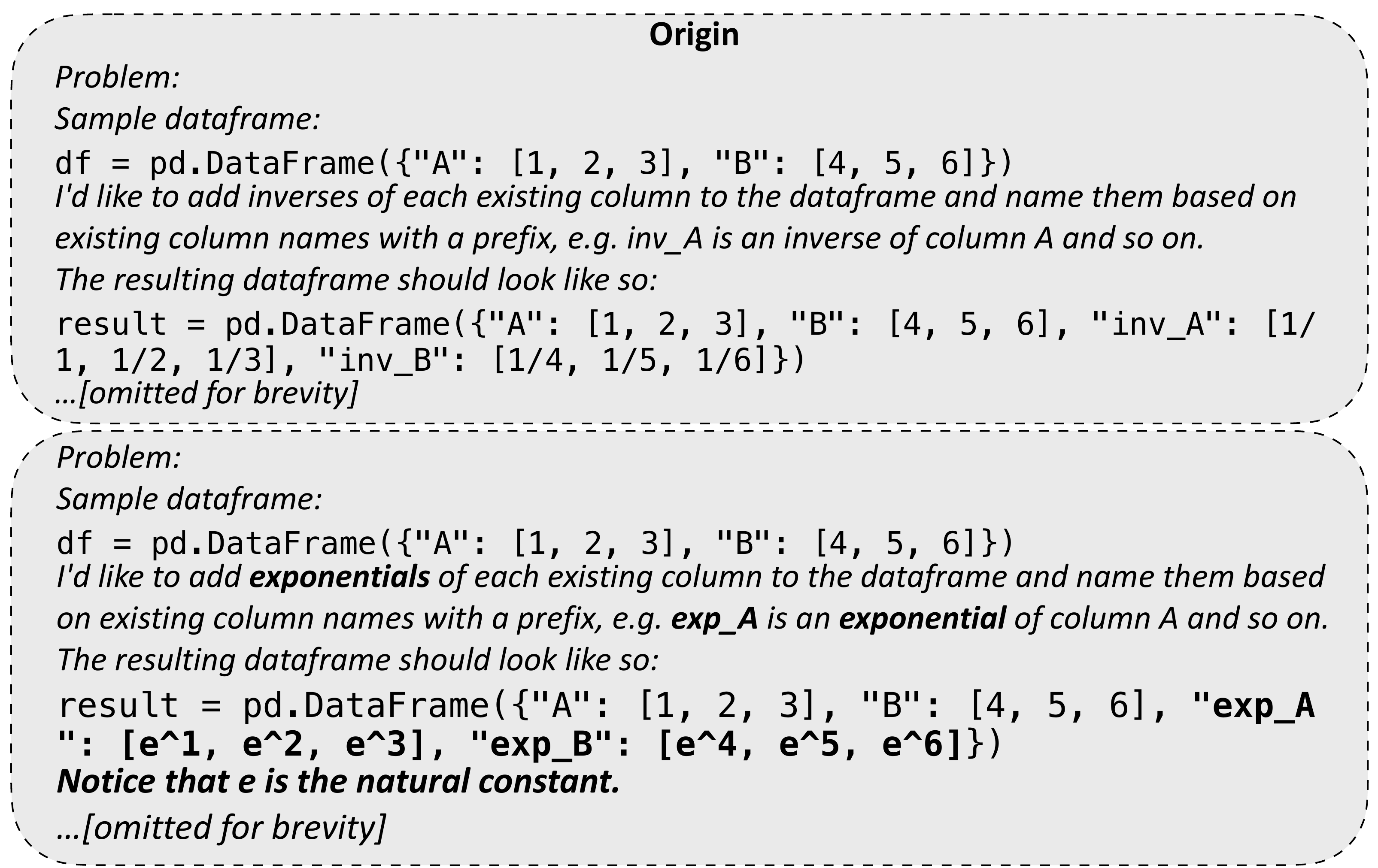}
    \caption{
    An example problem of semantic perturbation. ``inverse'' has been replaced with an analogy word ``exponential''.
    }
    \label{fig:example-A3}
\end{figure*}

\begin{figure*}[h]
    \centering
    \includegraphics[width=0.75\textwidth]{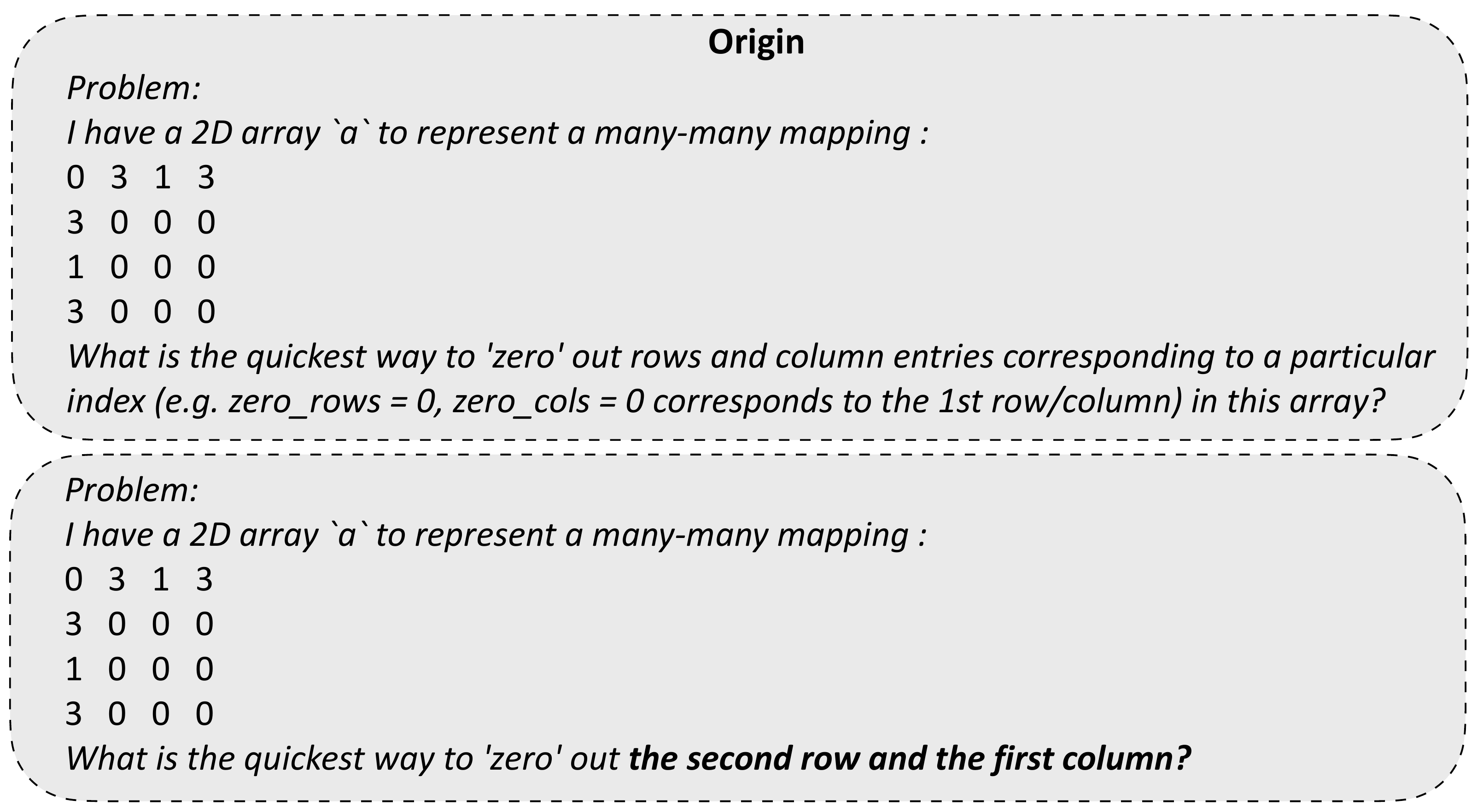}
    \caption{
    An example problem of semantic perturbation. The required index of rows and columns has been changed.
    }
    \label{fig:example-A4}
\end{figure*}

\begin{figure*}[h]
    \centering
    \includegraphics[width=0.75\textwidth]{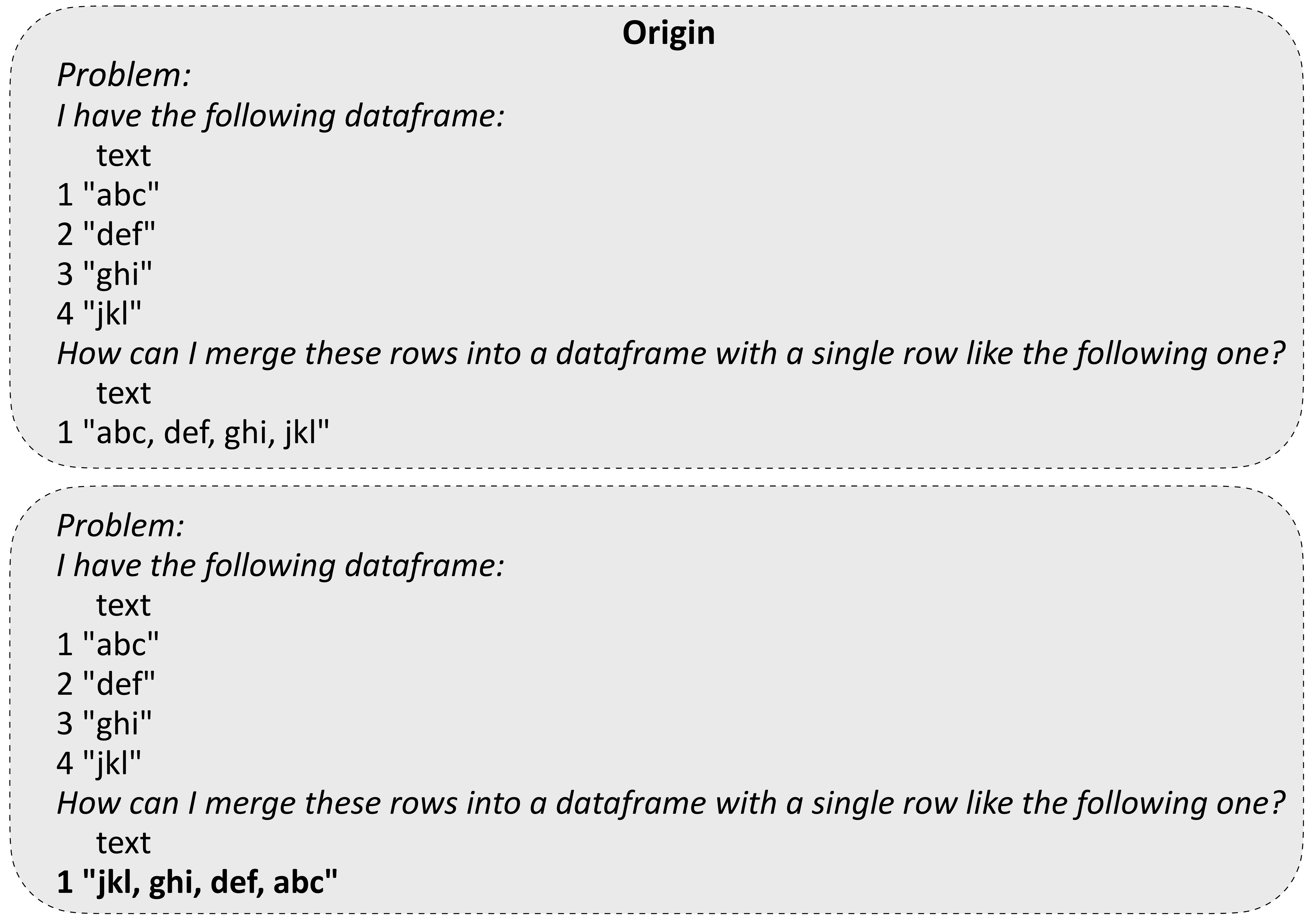}
    \caption{
    An example problem of semantic perturbation. The order of the desired string has been reversed.
    }
    \label{fig:example-A5}
\end{figure*}

\begin{figure*}[h]
    \centering
    \includegraphics[width=0.9\textwidth]{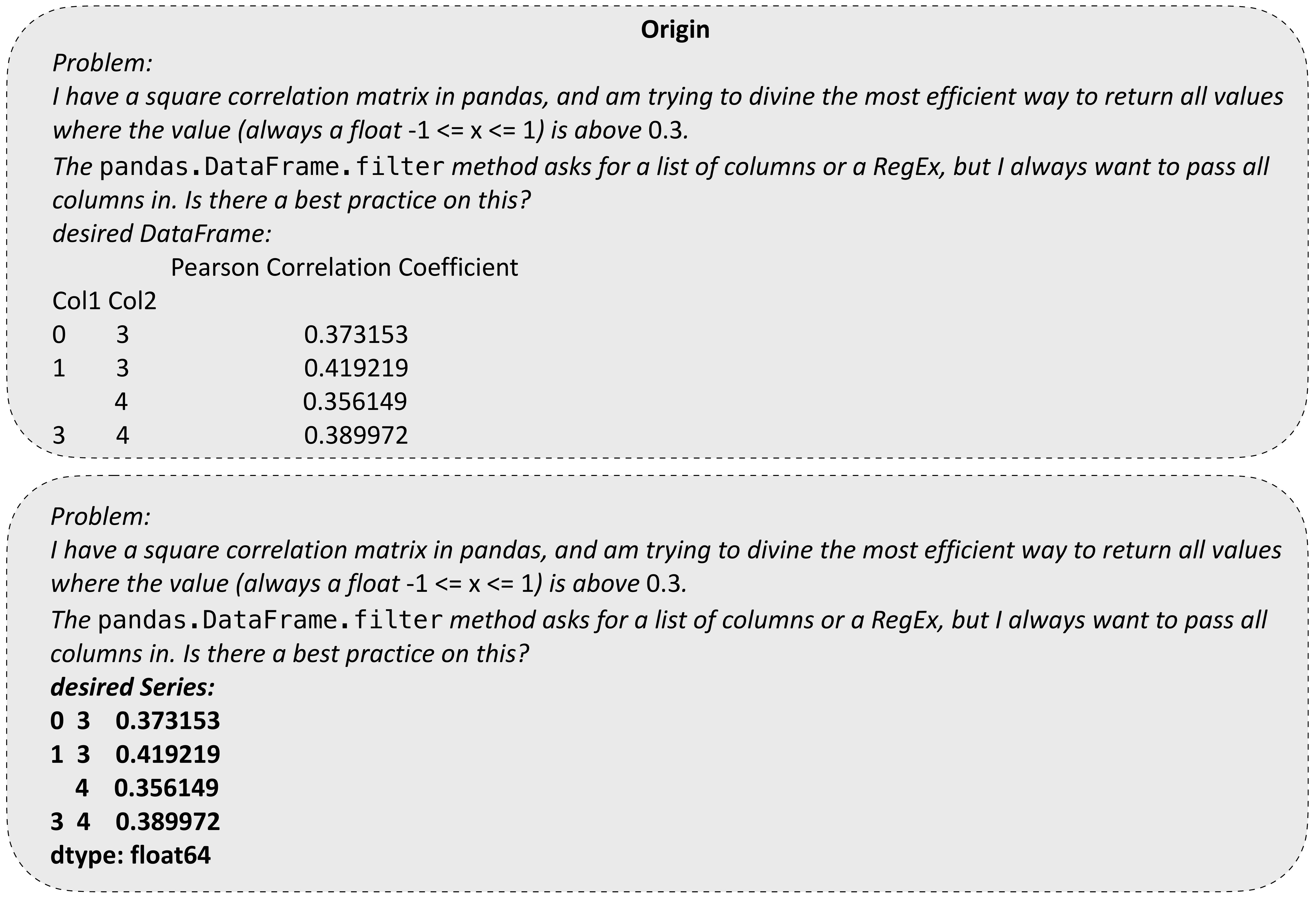}
    \caption{
    An example problem of semantic perturbation. The type of the desired result has been changed but the content still keeps the same.
    }
    \label{fig:example-A6}
\end{figure*}

Figure~\ref{fig:example-A3} shows how we replace keywords with analogy words in a \code{\pd} problem. The perturbed problem asks for applying an exponential function to column \code{A} and \code{B}. The problem in Figure~\ref{fig:example-A4} concentrates on changing the required index. Here we specify the target index on which to operate using ordinal numbers. Figure~\ref{fig:example-A5} gives an example of reversing the order. The desired output string is reversed(from ``abc,def,ghi,jkl'' to ``jkl,ghi,def,abc''). We expect the models to capture the information and handle the perturbation. Figure~\ref{fig:example-A6} shows an example of changing the type of the required result. Here we change the type from \code{pd.DataFrame} to \code{pd.Series}.

\begin{figure*}[h]
    \centering
    \includegraphics[width=0.9\textwidth]{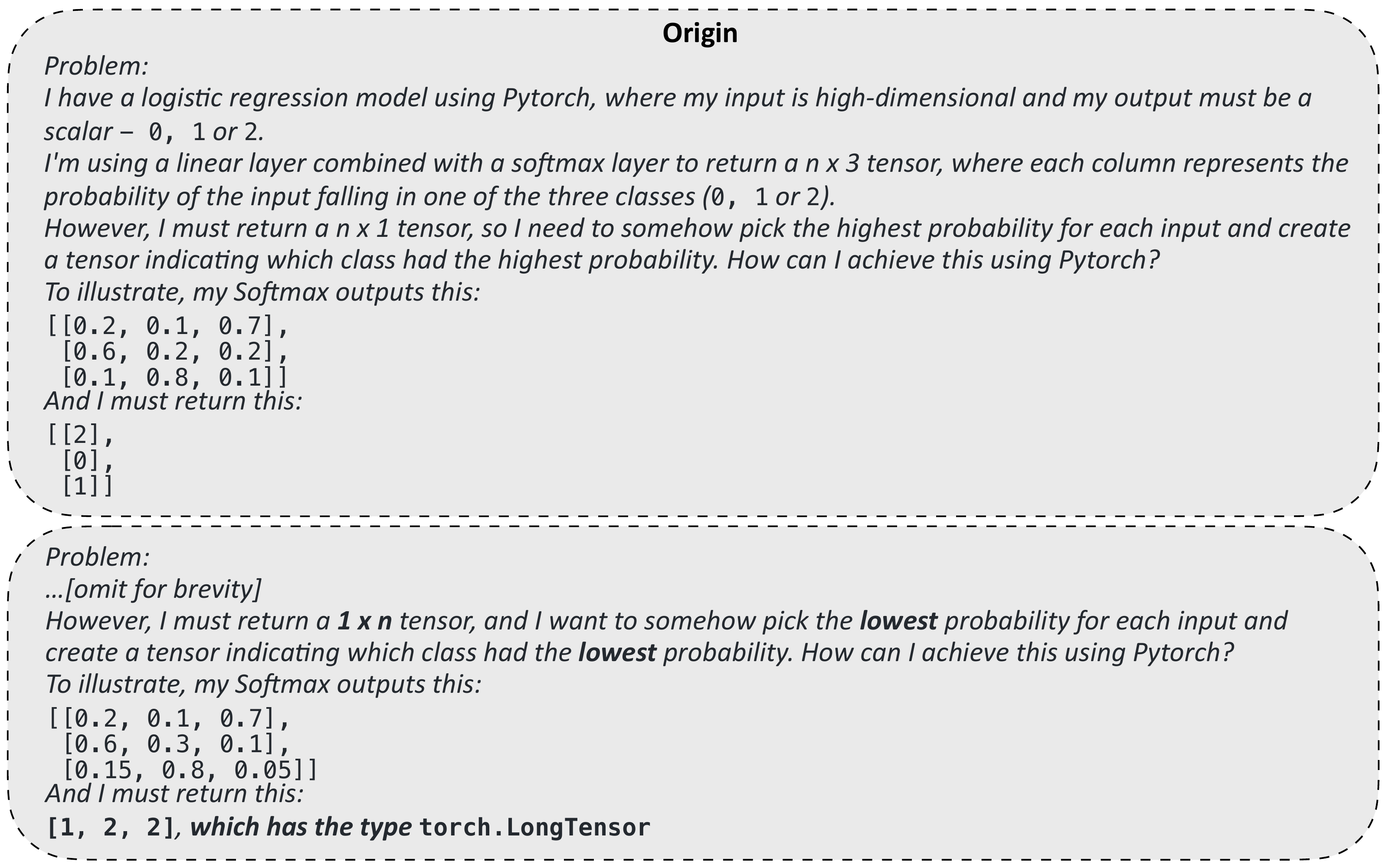}
    \caption{
    An example problem that is difficult re-written with a combination of surface and semantic perturbations
    }
    \label{fig:example-P1}
\end{figure*}

\begin{figure*}[h]
    \centering
    \includegraphics[width=1\textwidth]{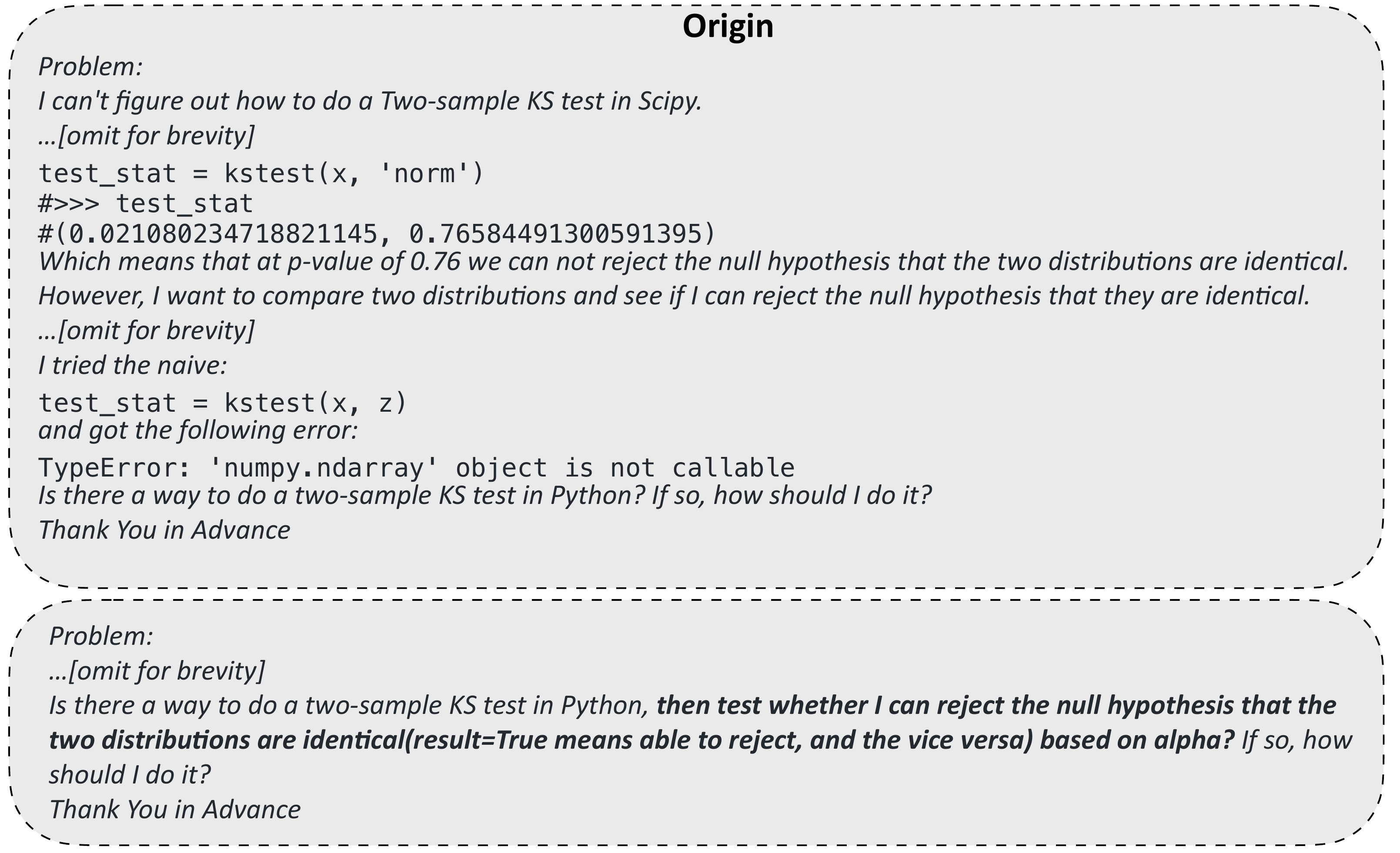}
    \caption{
    An example problem that is difficult re-written for more complexity
    }
    \label{fig:example-P2}
\end{figure*}

Figure~\ref{fig:example-P1} and Figure~\ref{fig:example-P2} demonstrate how we get difficult rewrites. The example in Figure~\ref{fig:example-P1} replaces ``highest'' with ``lowest'' and changes the shape of the desired output (from \code{n $\times$ 1} to \code{1 $\times$ n}). The example in Figure~\ref{fig:example-P2}, on the other hand, focuses on digging more perturbations that could increase the difficulty. The models should not only learn how to use a two-sample KS test but also learn how to interpret the result of the KS test.

\subsection{Prompt Format}
\label{sec:prompt-format-appendix}

\begin{figure*}[h]
    \centering
    \includegraphics[width=0.75\textwidth]{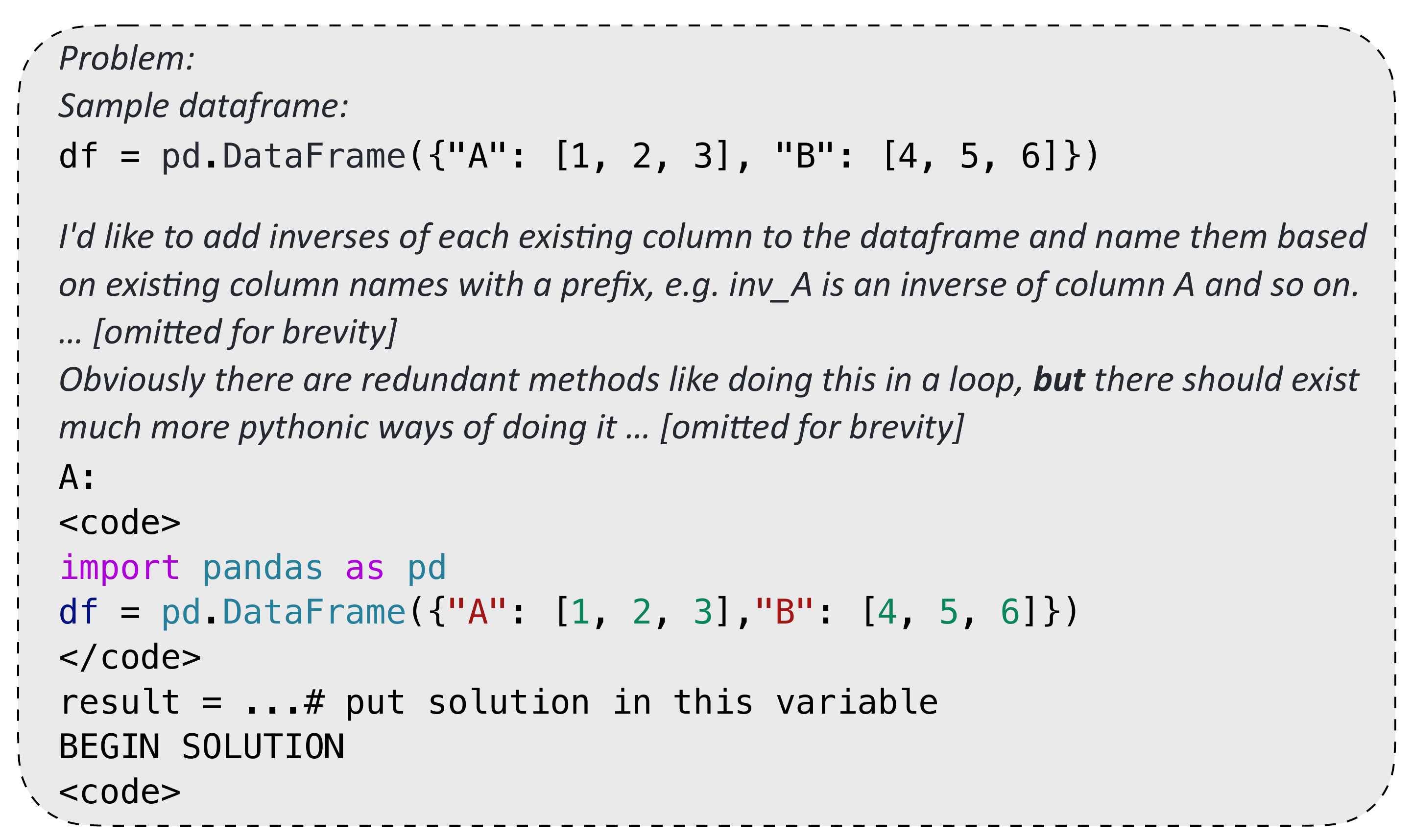}
    \caption{
    Completion prompt corresponding to \autoref{fig:example-datapoint}.
    }
    \label{fig:example-completion}
\end{figure*}

\begin{figure*}[h]
    \centering
    \includegraphics[width=1\textwidth]{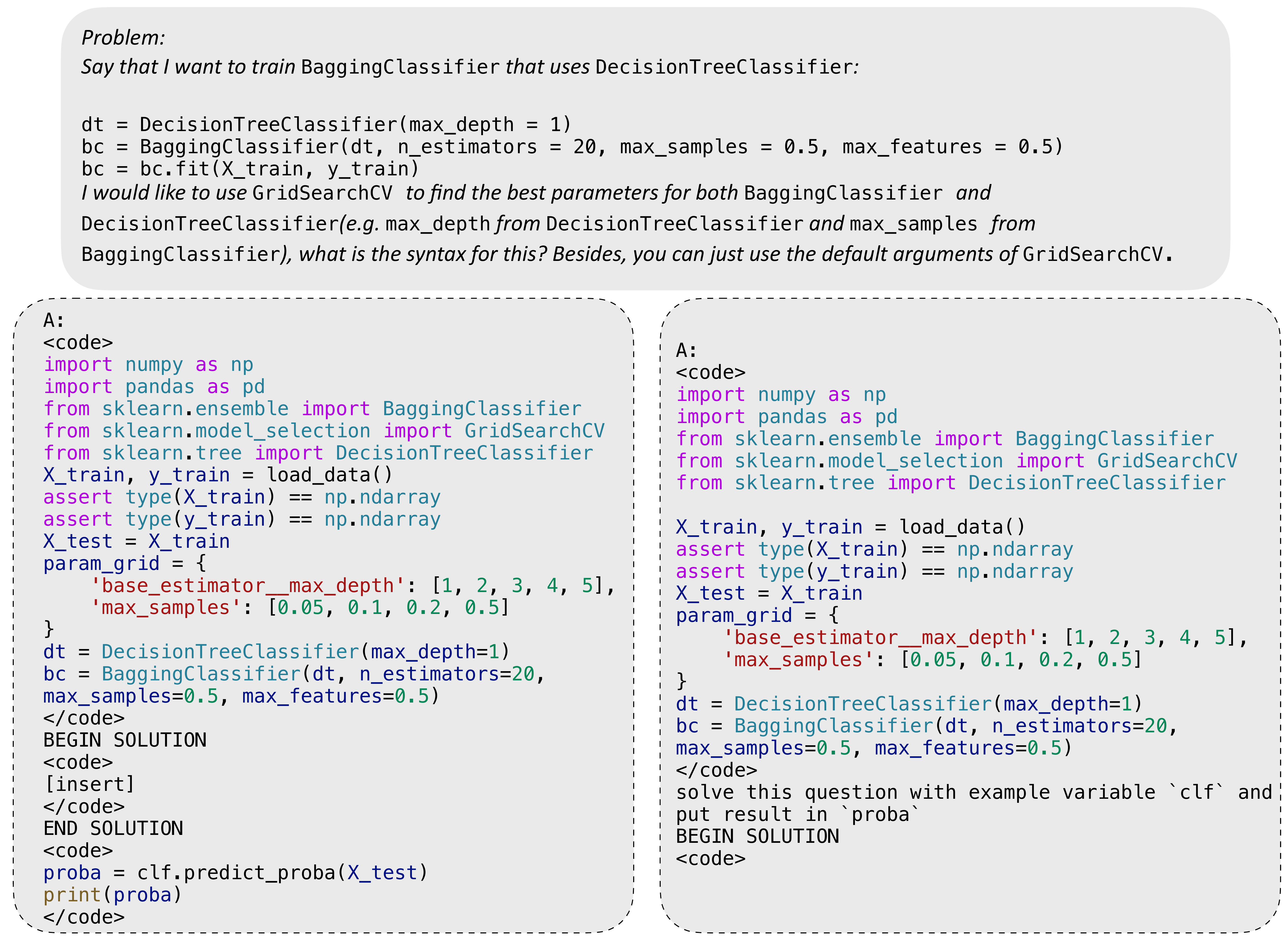}
    \caption{
    More complex Completion (on the right) prompt that requires additional information for a solution.
    }
    \label{fig:example-complexcompletion}
\end{figure*}

As we've mentioned in \Cref{sec:prompt-format}, we also provide a prompt of Completion format. Here are two examples (\Cref{fig:example-completion} and \Cref{fig:example-complexcompletion}) showing that we have to translate the code in the right context into natural language instructions as complementary information.

\section{Details of Experiments on \npHundred}
\label{sec:np100-appendix}
\npHundred is a collection of 100 \code{\np} exercises from \code{\np} mailing list, \SO, and \code{\np} documentation, which has been forked over 4.7k times on \github.

\begin{figure}[h]
    \centering
    \includegraphics[width=0.5\textwidth]{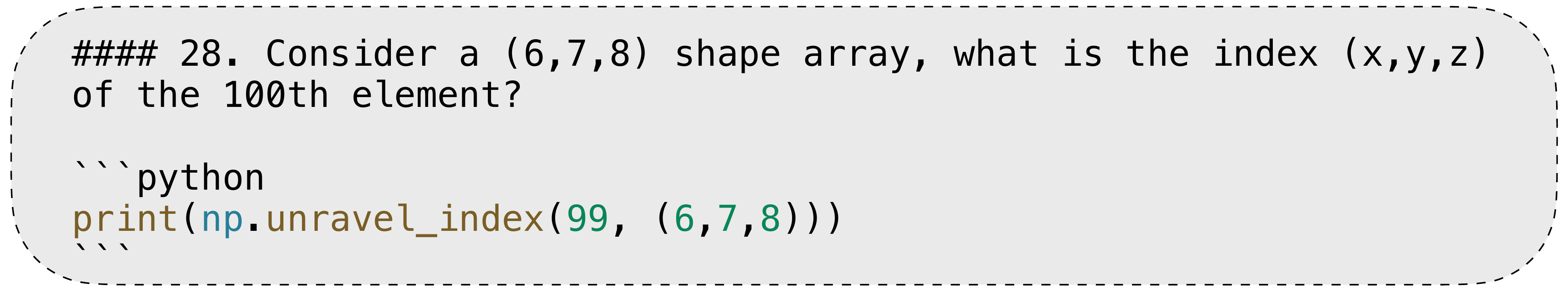}
    \caption{
    A \npHundred example.
    }
    \label{fig:numpy100-website}
\end{figure}

As shown in \Cref{fig:numpy100-website}, in the \npHundred problem set, each problem is given a short, one-sentence description with no code context, followed by a reference solution.

\begin{figure}[h]
    \centering
    \includegraphics[width=0.5\textwidth]{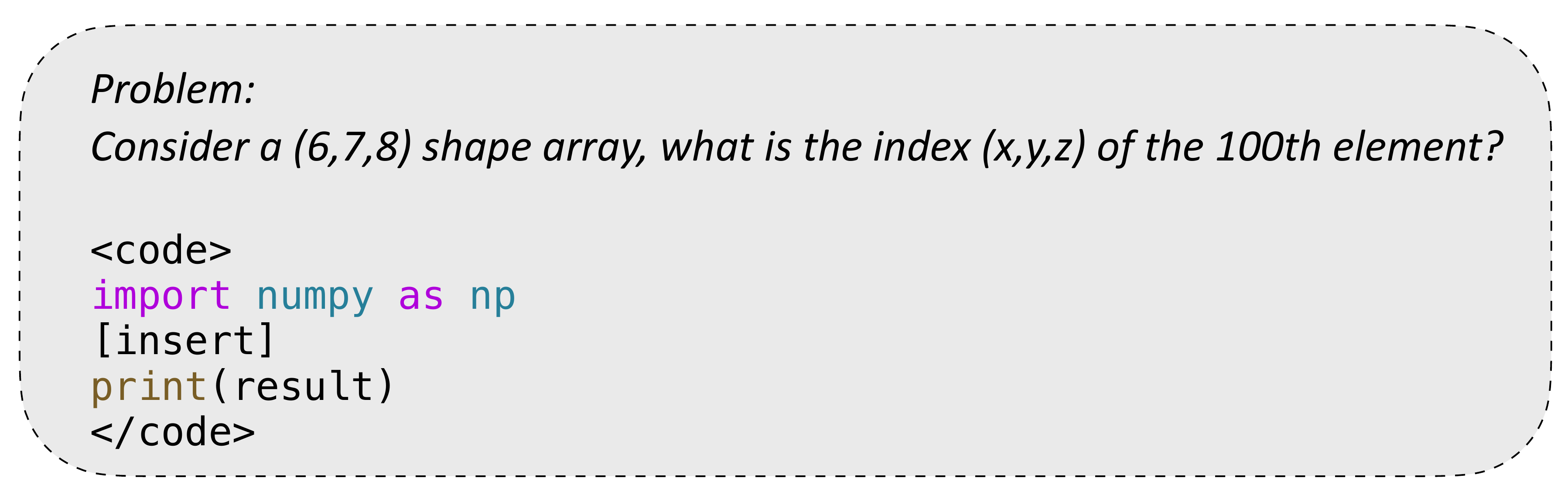}
    \caption{
    A \npHundred example prompt.
    }
    \label{fig:numpy100-origin}
\end{figure}

First, we wrote a code context for each problem and applied Insertion prompt, as shown in \Cref{fig:numpy100-origin}.

\begin{figure}[h]
    \centering
    \includegraphics[width=0.5\textwidth]{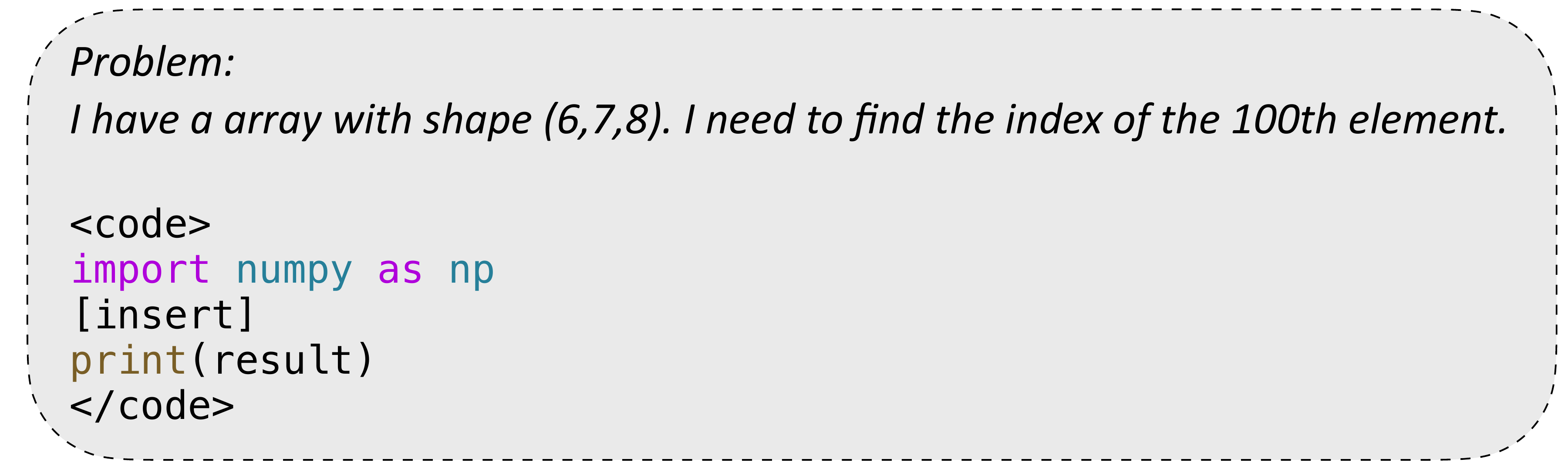}
    \caption{
    A \npHundred example of surface perturbation. We expressed the same description in different words.
    }
    \label{fig:numpy100-paraphrasing}
\end{figure}

\begin{figure}[h]
    \centering
    \includegraphics[width=0.5\textwidth]{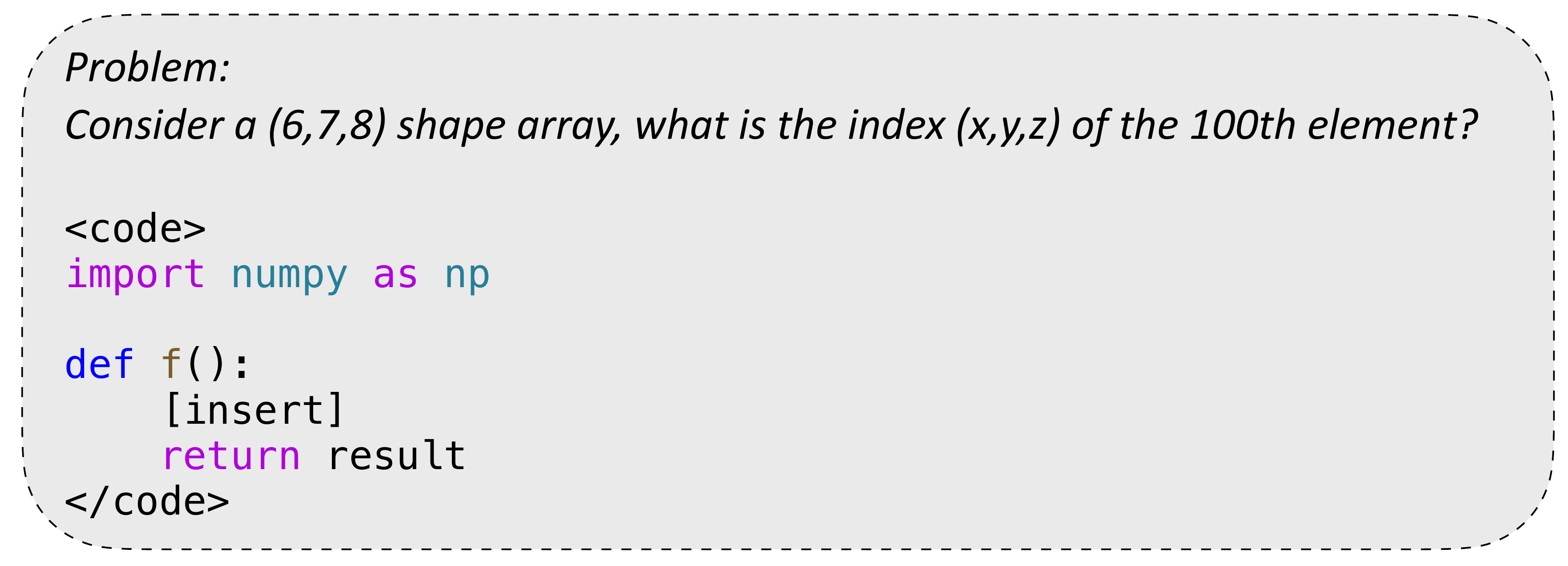}
    \caption{
    A \npHundred example of surface perturbation. We changed the code context.
    }
    \label{fig:numpy100-function}
\end{figure}

Then we paraphrased the problems and modified the code contexts as surface perturbations, as shown in \Cref{fig:numpy100-paraphrasing} and \cref{fig:numpy100-function}. 
We changed the description from ``Consider a (6,7,8) shape array, what is the index (x,y,z) of the 100th element?'' to ``I have an array with shape (6,7,8). I need to find the index of the 100th element.''. 
In another way, we changed the code context to require models to complete a given function.

\begin{figure}[b]
    \centering
    \includegraphics[width=0.5\textwidth]{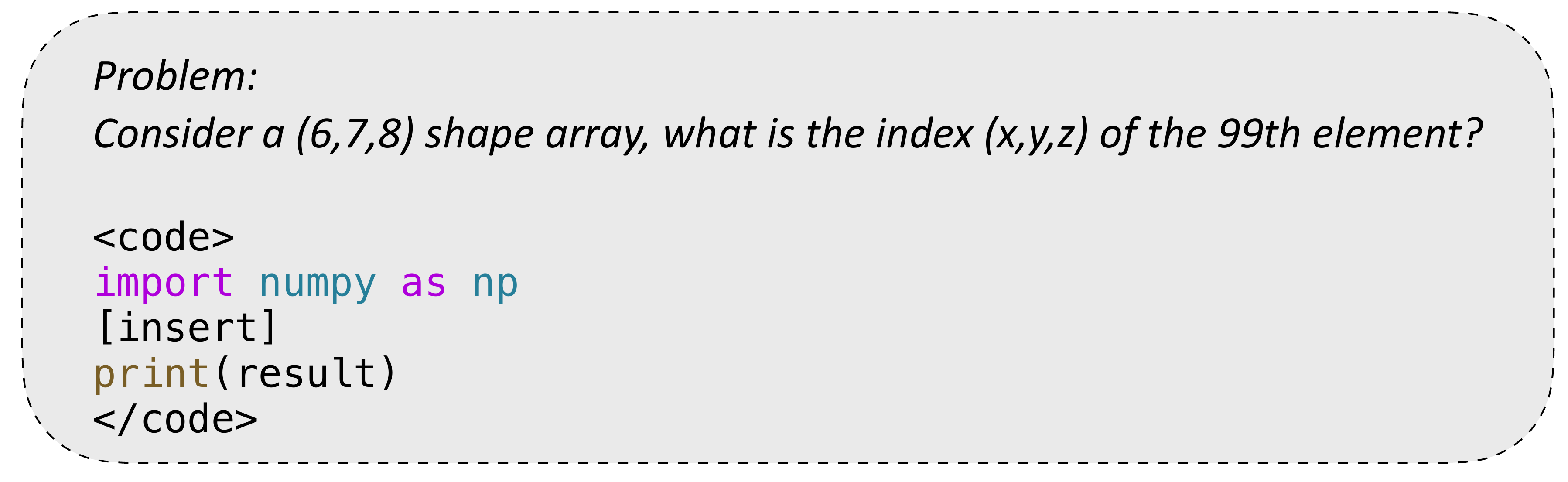}
    \caption{
    A \npHundred example of semantic perturbation. We only changed the required index.
    }
    \label{fig:numpy100-semantic}
\end{figure}

For semantic perturbation, we changed the requirements of the problems and also the semantics of the reference solutions without changing their difficulty. 
As shown in \Cref{fig:numpy100-semantic}, we changed ``100'' to ``99''.

At last, we equipped each problem and its perturbation with one test case and an automatic evaluation. Then we tested the performance of \codexTwo on them.
We sampled 20 problems from \npHundred and generated $10$ samples for each problem with temperature set to $0.7$, and top-p cutoff set to $0.95$.

\section{Error Analysis}
\label{sec:error}
\begin{figure*}[t]
    \centering
    \includegraphics[width=0.8\textwidth]{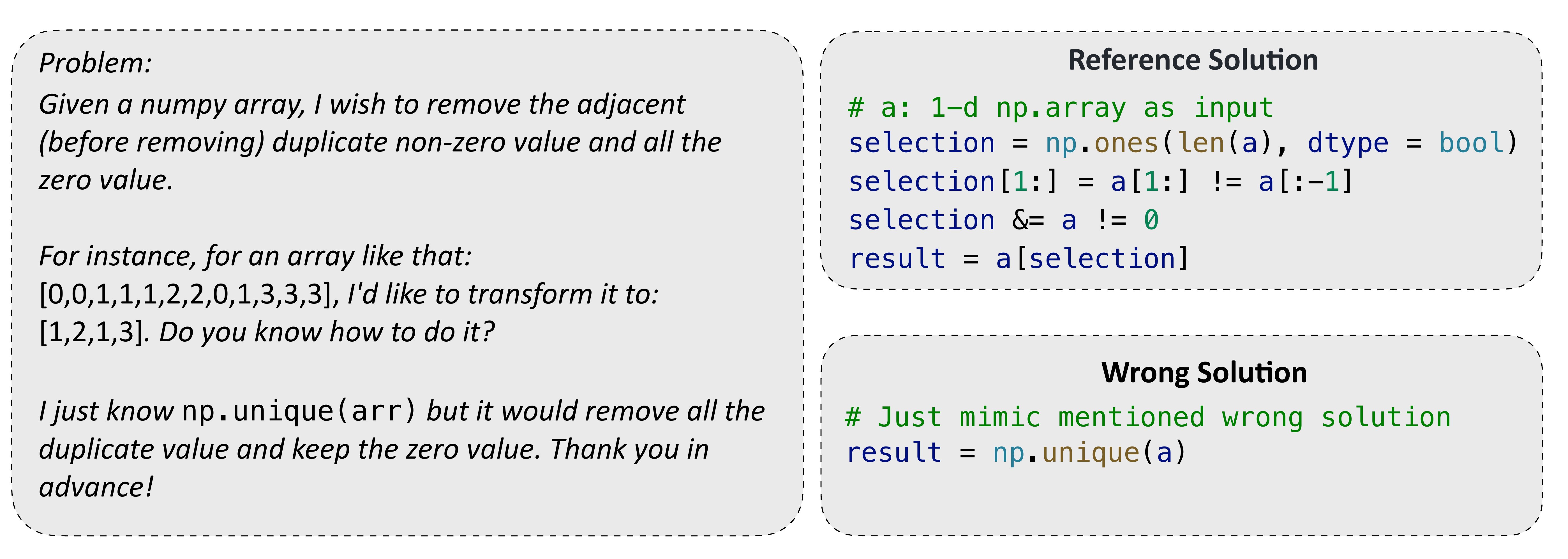}
    \caption{An example model mistake. The problem specifies a composite requirement, removing adjacent non-zero duplicates, which cannot be solved by a single operation.
    The model mistakenly generates a single operation that removes all duplicates.
    }
    \label{fig:compositionality-error}
\end{figure*}

\begin{figure*}[]
    \centering
    \includegraphics[width=0.9\textwidth]{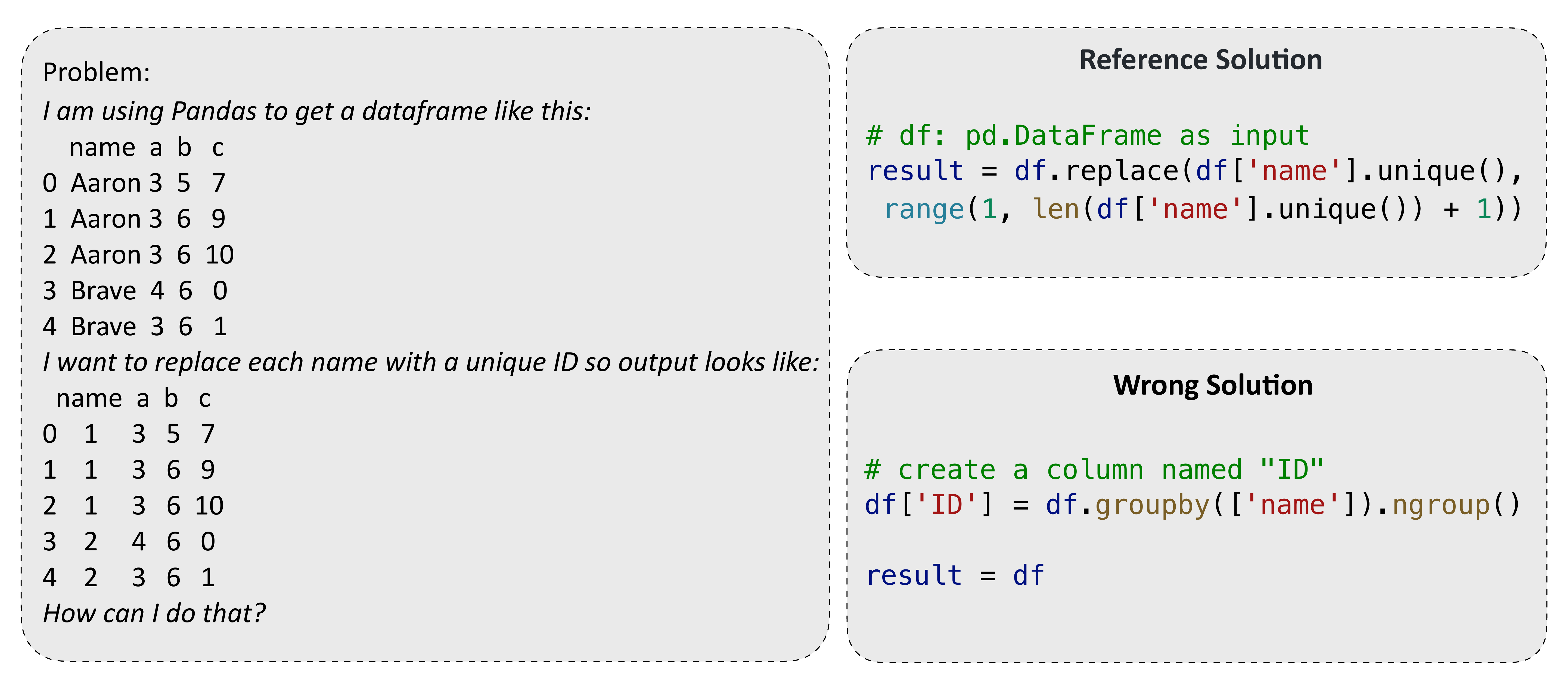}
    \caption{An example wrong solution that misunderstands
    the requirements and modifies on the wrong column.
    }
    \label{fig:col-error}
\end{figure*}

\begin{figure*}[]
    \centering
    \includegraphics[width=0.9\textwidth]{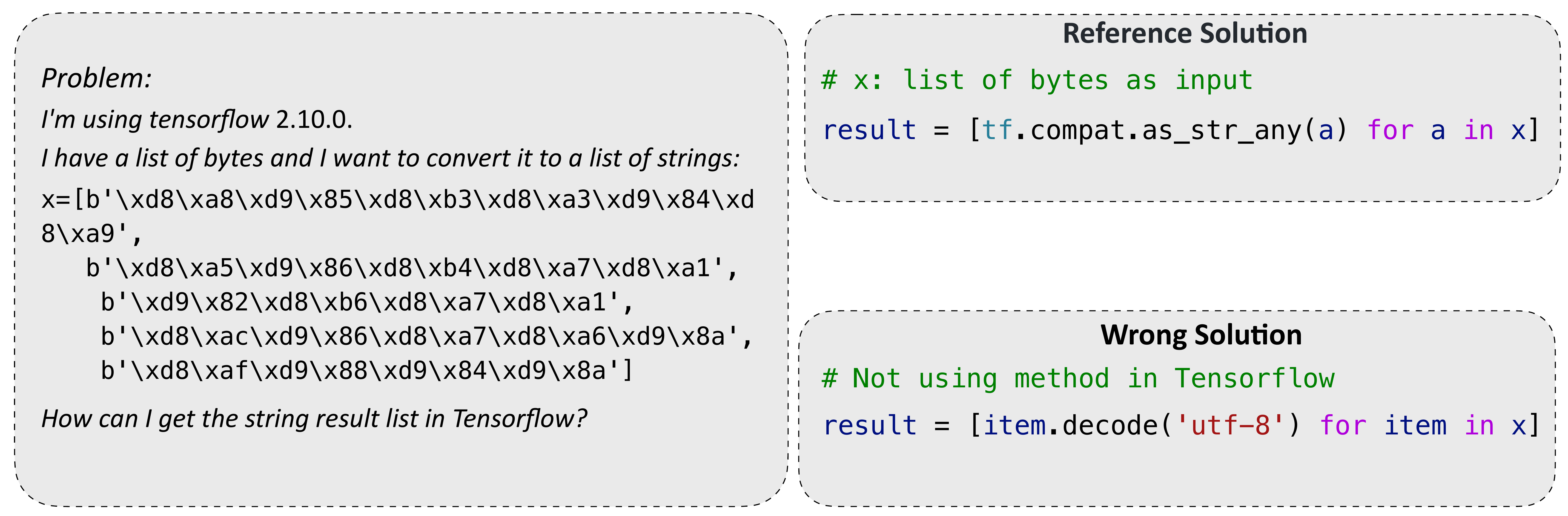}
    \caption{An example wrong solution that uses a common function instead of a function of \code{\tf}.
    }
    \label{fig:function-error}
\end{figure*}

We provide a preliminary error analysis by showing an example model error in Figure  \ref{fig:compositionality-error} and provide additional examples in Figure~\ref{fig:col-error} and \ref{fig:function-error}.
In this example, the problem asks for removing adjacent duplicated non-zero values in a given array, which cannot be satisfied by a single \code{\np} operation.
The reference implements this problem by creating a binary array representing the selection and performing two operations to meet the problem requirement.
However, we see \codexTwo fails on the composite request and attempts to answer the problem with a single method, \texttt{np.unique}, pointed out as incorrect in the problem already.. 
This example error demonstrates the challenges in \dscg problems, which require both natural language understanding and code generation abilities.

\clearpage

\end{appendices}

%%%%%%%%%%%%%%%%%%%%%%%%%%%%%%%%%%%%%%%%%%%%%%%%%%%%%%%%%%%%%%%%%%%%%%%%%%%%%%%
%%%%%%%%%%%%%%%%%%%%%%%%%%%%%%%%%%%%%%%%%%%%%%%%%%%%%%%%%%%%%%%%%%%%%%%%%%%%%%%
% APPENDIX
%%%%%%%%%%%%%%%%%%%%%%%%%%%%%%%%%%%%%%%%%%%%%%%%%%%%%%%%%%%%%%%%%%%%%%%%%%%%%%%
%%%%%%%%%%%%%%%%%%%%%%%%%%%%%%%%%%%%%%%%%%%%%%%%%%%%%%%%%%%%%%%%%%%%%%%%%%%%%%%
% \newpage
% \appendix
% \onecolumn

% You can have as much text here as you want. The main body must be at most $8$ pages long.
% For the final version, one more page can be added.
% If you want, you can use an appendix like this one, even using the one-column format.
%%%%%%%%%%%%%%%%%%%%%%%%%%%%%%%%%%%%%%%%%%%%%%%%%%%%%%%%%%%%%%%%%%%%%%%%%%%%%%%
%%%%%%%%%%%%%%%%%%%%%%%%%%%%%%%%%%%%%%%%%%%%%%%%%%%%%%%%%%%%%%%%%%%%%%%%%%%%%%%

\end{document}